\documentclass[conference]{IEEEtran}

\usepackage{cite}
\usepackage{amsmath}
\usepackage{amssymb}
\usepackage{amsfonts}
\usepackage{pifont}
\usepackage{graphicx}
\usepackage{datetime}
\usepackage{xcolor}
\usepackage{xspace}
\usepackage{listings}
\usepackage{booktabs}
\usepackage{extarrows}
\newtheorem{theorem}{Theorem}
\newtheorem{definition}{Definition} %[chapter]
\newtheorem{remark}{Remark}
\newcommand{\hide}[1]{}

\newcommand{\Major}{\texttt{Major}\xspace}
\newcommand{\Minor}{\texttt{Minor}\xspace}
\newcommand{\Patch}{\texttt{Patch}\xspace}
\newcommand{\version}{v}
\newcommand{\versions}{\mathcal{\set{\version}}}
\newcommand{\N}{\mathbb{N}}
\newcommand{\major}{\version_{(0)}}
\newcommand{\minor}{\version_{(1)}}
\newcommand{\patch}{\version_{(2)}}
\newcommand{\RSI}{RSI\xspace}
\newcommand{\RSIlong}{Remote Software Identification\xspace}
\newcommand{\fingerprinttechnique}[1]{#1\xspace}
\newcommand{\ftversionclaim}{\fingerprinttechnique{Version Claim}}
\newcommand{\ftfilestructure}{\fingerprinttechnique{File Structure/Content}}
\newcommand{\ftlistplugins}{\fingerprinttechnique{Sub-Software}}%List of Incorporated Software
\newcommand{\fthashdigest}{\fingerprinttechnique{File Hash Digest}}
\newcommand{\ftversionapi}{\fingerprinttechnique{Version API}}
\newcommand{\ftsecurityexploit}{\fingerprinttechnique{Security Exploit}}
\newcommand{\fterrorcode}{\fingerprinttechnique{Error Code}}
\newcommand{\ftfunctionoutput}{\fingerprinttechnique{Static Function Output}}
\newcommand{\ftversionspecific}{\fingerprinttechnique{Version Specific Features}}
\newcommand{\ftrequestresponse}{\fingerprinttechnique{Dynamic Request Response}}
\newcommand{\fterrorhandling}{\fingerprinttechnique{Error Handling}}
\newcommand{\ftsecurityexploitnd}{\fingerprinttechnique{Non-Destructive Security Exploit}}

\newcommand{\ftclassone}{Class 1: Version Label-Depending Techniques}
\newcommand{\ftclasstwo}{Class 2: Output Function-Depending Techniques}
\newcommand{\ftclassthree}{Class 3: Transition Function-Depending Techniques}
\newcommand{\ftclassfour}{Class 4: Security Flaw-Depending Techniques}
\newcommand{\ftclassfive}{Class 5: Dynamic Behavior-Depending Techniques}
\newcommand{\code}[1]{\texttt{#1}\xspace}
\newcommand{\fsm}{M}
\newcommand{\fsmset}{\mathcal{\fsm}}
\newcommand{\fsmtarget}{\fsm_\mathrm{trg}}
\newcommand{\fsmsource}{\fsm_\mathrm{src}}
\newcommand{\fsmchallenge}{\fsm_\mathrm{chl}}
\newcommand{\set}[1]{\MakeUppercase{#1}}
\newcommand{\state}{z}
\newcommand{\states}{\mathcal{\set{\state}}}
\newcommand{\startstate}{{\state_0}}
\newcommand{\inputalphabet}{\Sigma}
\newcommand{\inp}{\sigma}
\newcommand{\inputs}{\vec{\inp}}
\newcommand{\inputset}{\set{e}}
\newcommand{\outputalphabet}{\Gamma}
\newcommand{\outp}{\gamma}
\newcommand{\outputs}{\vec{\outp}}
\newcommand{\fout}{\omega}
\newcommand{\foutM}[1]{\fout_{#1}}
\newcommand{\ftrans}{\delta}
\newcommand{\party}[1]{\ensuremath{\mathrm{#1}}\xspace}
\newcommand{\prover}{\party{P}}
\newcommand{\provider}{\prover}
\newcommand{\verifier}{\party{V}}
\newcommand{\auditor}{\party{A}}
\newcommand{\user}{\party{U}}
\newcommand{\produce}{\ensuremath{\mathsf{Produce}}\xspace}
\renewcommand{\Pr}[1]{\mathrm{Pr}\left[#1\right]}
\newcommand{\secparam}{\lambda}
\newcommand{\protocol}[1]{\Pi_\mathrm{#1}}
\newcommand{\rsip}{\protocol{RSI}}
\newcommand{\rfpp}{\protocol{RFP}}
\newcommand{\decision}{\delta}
\newcommand{\decisionlog}{\Delta}
\newcommand{\true}{\texttt{true}\xspace}
\newcommand{\false}{\texttt{false}\xspace}
\newcommand{\distseq}{\mathsf{DS}}
\newcommand{\RFPlong}{Reverse Fingerprinting\xspace}
\newcommand{\RFP}{RFP\xspace}
\newcommand{\ORFP}{ORFP\xspace}
\newcommand{\candidateset}{\set{C}}
\newcommand{\func}{f}
\newcommand{\funcset}{\set{\func}}
\newcommand{\funcsethard}{\funcset^\star}
\newcommand{\funcsetgood}{\mathcal{\funcset}^\star}
\newcommand{\fgood}{\funcsetgood_\mathrm{I}}
\newcommand{\fbad}{\funcsetgood_\mathrm{B}}
\newcommand{\fugly}{\funcsetgood_\mathrm{D}}
\newcommand{\bad}[1]{\tilde{#1}}
\newcommand{\sw}{S}
\newcommand{\swfamily}{\mathcal{\sw}}
\newcommand{\swall}{\bad{\swfamily}}
\newcommand{\swchallenge}{{\sw_\mathrm{chl}}}
\newcommand{\swsource}{{\sw_\mathrm{src}}}
\newcommand{\swtarget}{{\sw_\mathrm{trg}}}
\newcommand{\negl}{\nu}
\newcommand{\dshard}{\mathsf{DSH}}
\newcommand{\interface}{I}
\newcommand{\interfaces}{\mathcal{\set{\interface}}}
\newcommand{\interfacesall}{\interfaces^*}
\newcommand{\database}{D}
\newcommand{\databases}{\mathcal{\set{\database}}}
\newcommand{\dbentry}[1]{\database[#1]}
\newcommand{\strategy}{\theta}
\newcommand{\strategies}{\Theta}
\newcommand{\randomness}{\phi}
\newcommand{\randsource}{\Phi}
\newcommand{\scheme}[1]{\textsf{#1}}
\newcommand{\rfps}{\scheme{R\kern0.065em F\kern0.065em P}}
\newcommand{\process}[1]{\mathsf{#1}}
\newcommand{\procsetup}{\process{Setup}}
\newcommand{\proctest}{\process{Test}}
\newcommand{\procoutput}{\process{Output}}
\newcommand{\timec}{t}
\newcommand{\forAll}{~\forall~}
\newcommand{\challenge}{c}
\newcommand{\expresponse}{e}
\newcommand{\response}{e'}
\newcommand{\epsilonrsi}{\varepsilon_0}
\newcommand{\epsilonrfp}{\varepsilon_1}
\newcommand{\cmark}{\ding{51}}
\newcommand{\xmark}{\ding{55}}
\newcommand{\auditlog}{\Lambda}
\newcommand{\signature}[2]{\operatorname{Sign}_{#1}(#2)}
\newcommand{\timenow}{t}
\newcommand{\prulelong}{\hspace*{2mm}\rule[2pt]{12mm}{0.3pt}\hspace*{2mm}}
\renewcommand{\prulelong}{}
\newcommand{\pruleshort}{\hspace*{2mm}\rule[2pt]{6mm}{0.3pt}\hspace*{2mm}}
\renewcommand{\pruleshort}{}
\newcommand{\pto}[1]{\xlongrightarrow[\hspace*{25mm}]{#1}}
\newcommand{\pgets}[1]{\xlongleftarrow[\hspace*{25mm}]{#1}}

\begin{document}
\title{Reverse Fingerprinting}

\author{\IEEEauthorblockN{Christian A. Gorke}
\IEEEauthorblockA{University of Mannheim\\
Mannheim, Germany\\
mail@christiangorke.de}
\and
\IEEEauthorblockN{Frederik Armknecht}
\IEEEauthorblockA{University of Mannheim\\
Mannheim, Germany\\
armknecht@uni-mannheim.de}}

\maketitle

\begin{abstract}
Software connected to the Internet is an attractive target for attackers: as soon as a security flaw is known, services may be taken under attack.
In contrast, software developers release updates to add further features and fix flaws in order to increase its security.
Consequently, a user of the software wants to have the latest secure version running.
However, if the software is provided as a service, e.g., as part of the cloud, the user relies on the service provider (SP) to perform such updates.
But when asking for the software version, the user has to trust the output of SP or his software.
Latter may be malformed, since updating software costs time and money, i.e., in comparison to changing a (false) version string.

The question rises how a client of a software service can provably determine the real software version of the running service at the SP, also known as \RSIlong (\RSI).
While existing tools provide an answer, they can be tricked by the service to output any forged string because they rely on the information handed directly by the SP.
We solve the problem of \RSI by introducing \RFPlong (\RFP), a novel challenge-response scheme which employs the evaluation of inherit functions of software versions depending on certain inputs.
That is, \RFP does not rely on version number APIs but employs a database consisting of function inputs and according outputs and combines them with a strategy and a randomness source to provably determine the version number.
%As an advantage, service typical interfaces are enough to determine the software version that is really running.
We also provide a theoretical framework for \RSI and \RFP, and describe how to create databases and strategies.
Additionally, \RFP can be securely outsourced to a third party, called the auditor, to take away the burden of the user while respecting liability.
We also provide an implementation and API to perform \RFP in practice, showing that most of the providers have installed the latest versions.
\end{abstract}


\section{Introduction}
%COMPUTERS
Computers are present in various forms in our everyday life, for example desktop PCs, laptops, servers, smartphones, watches, TVs, routers, surveillance cameras, or even cars.
%Modern cars are a computer with a lot of hardware on four wheels.
A computer can be seen as a system that consists of storage, computational power, and interfaces to send and receive messages.
Due to smart homes, smart grids, and the Internet of things, the amount of connected devices and computers will rise even more in the years to come.
%This may be a keyboard, mouse, monitor, or even the Internet.
%It is used to connect with other entities, such as humans or further computers, e.g., as part of a network.

%SOFTWARE
At the end, most computers rely on software to perform nearly any task which also enables them to communicate with each other, especially over a network like the Internet.
Therefore, from a connection point of view, any computer on the Internet is somehow connected to any other computer on the Internet.
This allows for attacks from around the world, resulting in security exploits such as data breaches, denial of service attacks, and credential phishing.
%While an attack may also be performed by someone who has direct contact to the device, the threat of being attacked does overwhelmingly come from the Internet.
In most cases, security faults exists due to insecure software, allowing attackers to perform actions they would not be able to do otherwise.
Services and applications which run on the (public) cloud have negligible downtime and hence are prone to attacks.
For example, authenticating to a service, reading and writing data, or performing computations and spreading to other systems.
One reason for this situation is that the software employed today is very complex, reaching up to millions of lines of code for a single program.
Hence, software is difficult to manage and verify at scale regarding their security.
On the other hand, software developers often provide updates and patches to enhance and further secure their products.
These should be installed as soon as possible to prevent attacks, especially on systems connected to the Internet.
Unfortunately, there is also software running which is not maintained anymore and the only option for a user to be secure may be to switch to another software.

%Two kinds of security flaws exist in this case:
%the ones that are known by the software developer and the ones that are not (yet) known.
%Former can be addressed by updating software and providing patches to fix the errors.
%However, installing updates requires some effort from the user, even if updating can be automatized.
%Security flaws that are not known by the software provider are called zero-day exploits, if an attacker takes advantage of them..
%This is very important, since security flaws will be exploited sooner than later first by big parties with a lot of resources, and after an attack is known, even script kiddies are able to attack the user's data or processes.
%This results in the loss of data, customers, trust, and introduces new threats like bot networks.
%It especially becomes bad, when software is employed that is no longer supported by the developer, i.e., functional and security updates are no longer provided.

%CLOUD
Since the advent of cloud computing, software is offered by cloud providers as a service (CSP, cloud service provider), for example Software at a Service (SaaS) or Platform as a Service (PaaS).
In those cases, the user does not administrate the software and relies on the provider to install updates.

%IN- & SECURE CLOUD SOFTWARE
Due to the advancements in browser technology, web services and applications have become a central part of computer usage and hence an attractive target for hackers, since they often are publicly accessible.
Services running on the Internet or a cloud cover a range of applications, for example the popular content management system (CMS) Wordpress \cite{wordpress}, the most used programming language on the web, PHP \cite{php} (about 79\% of all web pages\cite{phpusage}), databases like MySQL \cite{mysql}, or services used by the system such as SMTP\cite{rfc5321}.
To offer a fair usage, these services have to be available over the Internet all the time at any place.
This, however, attracts attackers while continuously new security flaws are discovered.
For example, the CVE database contains over 6000 vulnerabilities for PHP alone\cite{cvephp}.
Besides flaws, new versions with vastly improved security features are released that should be used over old versions.
Additionally, some versions like PHP 5 and 7.0 will be no longer supported and should be updated as soon as possible\cite{phpsupport}.
Distributed public services also must be updated to fix security flaws, such as Bitcoin to prevent direct financial loss\cite{bitcoinbug}.

In conclusion, updating software is a foundation of todays security and part of the Internet in general.
Due to the service oriented structure of the cloud updating becomes the task of the service provider, who, in turn, wants to minimize costs and avoid configuration and dependency changes.
If the provider is careless, there will be time windows for attacks.
To cover this problem, CSPs blank out the version information on the service, i.e., employing security by obfuscation.
But this makes it also difficult for the honest customer who may wonder if the most recent software version is running on the service, e.g., to run own applications on top.
Of course, the customer can query the provider or the software itself for the version information, but both may answer whatever the provider wants the customer to believe.
This is crucial since the customer thinks she gets the version number, while in reality she may be receiving an arbitrary string.
Obviously, such a CSP may violate the mutual agreed Service Level Agreements (SLA), but customer may need some time to detect this behavior.

%CORE GOAL with RELATED EXAMPLES
Ultimately, it comes down to the following question:
\textit{How can a customer of a remote software reliably and efficiently prove that the service provider is behaving correctly, i.e., runs software in the correct version?}
In other words, the customer's goal is to identify the running software, i.e., determine its version number.
Various methods cope with the problem of finding and detecting insecure software, mainly vulnerability scanner (e.g., Nessus\cite{nessus}) and penetration tests.
The first are looking for known security flaws (e.g., CVE) but require certain access roles or structural execution, such as execution on the CSPs machine by the CSP.
The latter search for exploits like insufficient parsing of input values.
While this can be performed by external parties, default interfaces are used to determine the version number of a software (e.g., by Metasploit\cite{metasploit} or Burp Suite\cite{burp}).
In fact, existing solutions do not provide a sufficient level of security and can easily be fooled by a malicious service provider (see Section~\ref{sec:sota}.

%SUMMARY SOLUTION & CONTRIBUTIONS
In this paper, we will present the \RFPlong (\RFP) scheme as an answer to the postulated question.
\RFP does not rely on versioning interfaces or the support of the CSP to determine the software version number.
Overall, we give the following contributions:
\begin{description}
\item[Formalization of Software and Versioning]\ \\
We formalize software in general as finite state machines (FSM), where each software is represented by a unique FSM with a version number as label.
This allows us to include software development properties, such as co-existent version numbers, i.e., branching, and deprecated software, which otherwise makes differentiating between any software version challenging.
\item[Fingerprinting Techniques and Classification]\ \\
We analyze existing software fingerprinting techniques and group them into classes.
We further show that each existing fingerprinting technique lacks security properties to guarantee determination of the software version number.
As a solution, we introduce three novel secure fingerprinting techniques.
\item[\RSIlong (\RSI)]\ \\
We give a formal framework for \RSI including system model, attacker model, and protocol.
\item[\RFPlong (\RFP)]\ \\
As the main contribution of this paper, we present \RFPlong, a novel scheme that can be employed as a building block for \RSI.
It enables auditing of the provider, provably yielding the actual installed software version.
\RFP leverages a challenge-response protocol that relies on intrinsic functions of each individual software version to distinguish between them.
A specifically created database holds all tests, which are applied using various strategies resulting in a sequence of tests.
Furthermore, \RFP randomizes challenges, verifies time constraints of responses, and is applicable over multiple interfaces and software types.
We also present extensions of \RFP, such as including an auditor to take away computational and storage burden of the customer.
\item[Formal Security Analysis]\ \\
To the best of our knowledge, we are the first providing a security analysis of fingerprinting techniques and an \RSI scheme, i.e., \RFP.
We show that \RFP is secure against CSP response caching and precomputation, proxy-forwarding of challenges, and erroneous halts, while determining reliably the remote software version.
\item[Implementation and Real-World-Applicability]\ \\
We provide insights in how to build a database for \RFP and give examples for strategies, which describe the order to perform software version dependent tests.
Additionally, we apply \RFP to real-world applications and show that \RFP increases the security overall.
\end{description}

\section{Real-World Situation}
We will now define software versioning, fingerprinting, and software identification as foundation for following sections.
Furthermore, we summarize state of the art fingerprinting techniques and demonstrate their shortcomings with a practical attack.

\subsection{Software Versioning}\label{sec:versioning}
For now, let software be an algorithm or a machine which is in some state, receives inputs, performs processing, changes to other states, and yields an output.
We give a formal definition of software in Section~\ref{sec:model}.
When a software reaches a certain state, it gets assigned a unique label.
\begin{definition}[Software Version and Software Family]
A \emph{software version} is a label, e.g., a text string or a number, assigned by the software developer that uniquely identifies a certain state of a software.
Each software belongs to a set of softwares, called a \emph{software family}, where each member of the set has the same software version ancestor.
\end{definition}
A software version label may consist of both a name in and a number, e.g. Windows Vista for Windows 6.0 Build 6000.
However, most modern software sticks to semantic versioning\cite{semanticversioning} for releases.
Briefly, a version number is formatted as \Major{}\texttt{.}\Minor{}\texttt{.}\Patch, where \Major, \Minor, and \Patch\ are numbers.
\Major is increased when incompatible API changes are made, \Minor when backwards-compatible functionalities are added, and \Patch when backwards-compatible bugs are fixed.
At the end, we are interested in the version number, not the label.

Let $\version := (\major, \minor, \patch)$ be a software version, where $\major,\minor,\patch$ are elements from a totally ordered set, e.g., $\N_0$, represented by \Major, \Minor, and \Patch.
Let $\versions$ be the set of all software versions belonging to the same software family.
For $\version, \version' \in\versions$ it holds
\begin{align*}
\version < \version' \quad\Longleftrightarrow\quad & \left(\major<\major'\right) \quad\text{or}\\
&\left(\major=\major' \land \minor<\minor'\right) \quad\text{or}\\
&\left(\major=\major' \land \minor=\minor' \land \patch<\patch'\right),
\end{align*}
otherwise we have $\version \geq \version'$.
Note that we can map any software version format to the given structure.

\subsection{Fingerprinting}\label{sec:fingerprinting}
%\begin{definition}[Fingerprinting]
The identity of an entity, e.g., a person or an object, can be represented by a unique (digital) \emph{fingerprint} of the entity itself.
If the identity is not given, a certain amount of information about the entity is required in order to still build such a fingerprint.
\emph{Fingerprinting} refers to reliably determining the fingerprint of an entity.
%\end{definition}

For example, the identity of a user is combined with gathering properties of unique characteristics in order to track the user across multiple websites and devices, in fact relying on browser configurations and behavior of the user to build the fingerprint \cite{DBLP:conf/sp/LaperdrixRB16,DBLP:conf/ndss/CaoLW17,evercookie}.
Another example are hash values computed over files -- also known as ``checksums''.
While fingerprinting a user yields the identity of the user, fingerprinting a software results in the software identity, i.e., the software version.

\subsection{Auditing and Software Identification}
An enormous variety of different software is used everyday nearly in every part of our life.
For example, software is employed on the Internet, the cloud, computers, smartphones, Internet of things, networks, storage, automation, industry, etc.
On top, for each use case often multiple software solutions exist, e.g., MySQL\cite{mysql} or PostgreSQL\cite{postgresql} for a relational database.
For nearly each software, new versions are being released continuously, containing new features and fixing security flaws of previous versions.

Since security flaws are discovered and published every day, updating the operating software is crucial to have the latest security updates installed.
Hackers and automated attacks regularly exploit outdated software vulnerabilities resulting in data breaches, data loss, or functional disability, to name a few.
Furthermore, a software may depend on other software with a certain software versions, implicitly relying on a correct determination of the version number in order to work properly.
Consequently, checking whether the latest software version is running or not is a standard task in almost all auditing processes.
%Since most of the software used is not developed by the customer himself, he will need to perform an audit of the software he is using to ensure the correct version is installed.
Auditing refers to performing a check on the software and to convince the verifying party, e.g., the customer, that the software provider is compliant to some kind of service level agreement (SLA), i.e., providing a software in a certain version.
In other words, auditing yields the correct software version of the running software, where correct refers to the most recent version or the version they agreed on in the SLA.
Hence, we need to perform fingerprinting of the software to perform this kind of audit, since this will result in the software version.
The process of software fingerprinting gives a solution to \emph{\RSIlong} (\RSI).

\subsection{Shortcomings of State of the Art Solutions}\label{sec:sota}
In order to solve the problem of \RSI, many different solutions have already been developed and implemented.
A description of different implementations and an overview of related work can be found in Section~\ref{sec:relatedwork}.
As mentioned before, fingerprinting can be achieved in many different ways, we call a method to obtain a fingerprint a \emph{fingerprint technique}.
For software, a fingerprint technique aims to determine the software version of the currently running software.
This is usually achieved by sending a request to the running software and evaluating its response to finally output the software version.

To the best of our knowledge, state of the art solutions employ one or more fingerprint techniques out of four different general types of fingerprinting techniques, denoted as class.
We now will assign all existing fingerprinting techniques to one of the four classes and elaborate class specific properties.
Each fingerprinting technique is described in detail in Appendix~\ref{sec:ftdesc}, the according subsection is given in parentheses below.

\begin{description}
\item[\ftclassone]\ \\
This class of fingerprinting techniques asks the audited software to yield a static output containing its version label, usually in form of a string, e.g., the footer of a website.
The determined software version is the returned label.

Fingerprinting techniques: \ftversionclaim (\ref{sec:ftversionclaim}), \ftversionapi (\ref{sec:ftversionapi}), \ftversionspecific (\ref{sec:ftversionspecific}), \ftlistplugins (\ref{sec:ftlistplugins}), and \fterrorcode (\ref{sec:fterrorcode}).
\item[\ftclasstwo]\ \\
This class of fingerprinting techniques asks the audited software to output certain static values, e.g., one or more files characteristic for a certain version number.
Next, theses values are evaluated, for example matched by regular expressions or used as input into a hash function (file checksum).
At the end, the result is compared to already known values computed beforehand which yields the determined software version.

Fingerprinting techniques: \ftfilestructure (\ref{sec:ftfilestructure}), \fthashdigest (\ref{sec:fthashdigest}), \ftlistplugins (\ref{sec:ftlistplugins}), and \fterrorcode (\ref{sec:fterrorcode}).
\item[\ftclassthree]\ \\
This class of fingerprinting techniques requires the audited software to perform certain functions that are available only for certain software versions in a static order.
Then, multiple requests are employed to deduce a range of software versions.

Fingerprinting technique: \ftfunctionoutput (\ref{sec:ftfunctionoutput}).
\item[\ftclassfour]\ \\
In this class, security flaws of certain software versions are exploited in order to determine the version number.
Note that if a security flaw leads to a halt of the software, e.g., crash or deadlock, no further tests can be performed.

Fingerprinting technique: \ftsecurityexploit (\ref{sec:ftsecurityexploit}).
\end{description}

Let us briefly discuss the security of these four classes in regard of reliably determining the software version.
A full security analysis of all four classes is given in Appendix~\ref{sec:ftanalysis}.
The first class is forwarding a value being output by the software without any further check, that is the verifier believes everything the software replies.
The second class is similar to the first one, with the difference that the output of the software is transformed in some form, e.g., by applying filters.
While both classes claim to be software version-dependent, the third class requires not only static values from the software, but also some form of behavior.
However, note that this behavior may be precomputed by a malicious software or may be easily\footnote{For example, an easy to fake function may return a certain configuration value introduced with a given software version.} malformed, too.
Finally, the fourth class is not a reliable choice for fingerprinting, because exploiting a security flaw that is not yet fixed in the installed software version (which may be the case for all but the newest) will (maybe unintentionally) crash the software.
Hence, the verifier gains not much information about the actual software version (maybe except that it is not the newest one).
To the best of our knowledge, class 3 is only done by one implementation \cite{sqlmap}, and class 4 is only used by malicious users to attack software.

At the end, no class is fulfilling the goal of \RSI if the software or the party running the software is malicious.
The reasons are mainly that these techniques are either relying on static values chosen by the provider, i.e., values that do not depend on the randomness of the verifier, or the reply can be precomputed or faked easily.

\subsection{Practical Attack on Fingerprinting Techniques}\label{sec:practicalattack}
We will now give an example on how to trick \RSI implementations that employ a Class~1 fingerprinting technique, such as \ftversionclaim.
As described above and shown in Appendix~\ref{sec:ftanalysis}, a Class~1 fingerprint technique is not secure since it relies on fixed strings chosen by the provider.

Let us assume the software provider runs a software that is used by a customer who is going to audit the software.
In this example, the software employed is PHP \cite{php} in version 7.1.1 running on the operating system Ubuntu \cite{ubuntu} in version 16.04.3.
We have chosen the programming language PHP, since about 79\% of all websites are processed by PHP\cite{phpusage}.
Next, the provider runs the bash script in Listing~\ref{lst:vcas} (cf. Appendix~\ref{sec:vcas}) to manipulate the version number of PHP to the value \emph{20.9.85-car}, adding even extra version info to \Patch.
Then, the PHP functions \code{phpinfo()} and \code{phpversion()} as well as any related function with the goal to output the softare version, will return \code{20.9.85-car}.

\begin{figure}
\centering
\includegraphics[width=\columnwidth]{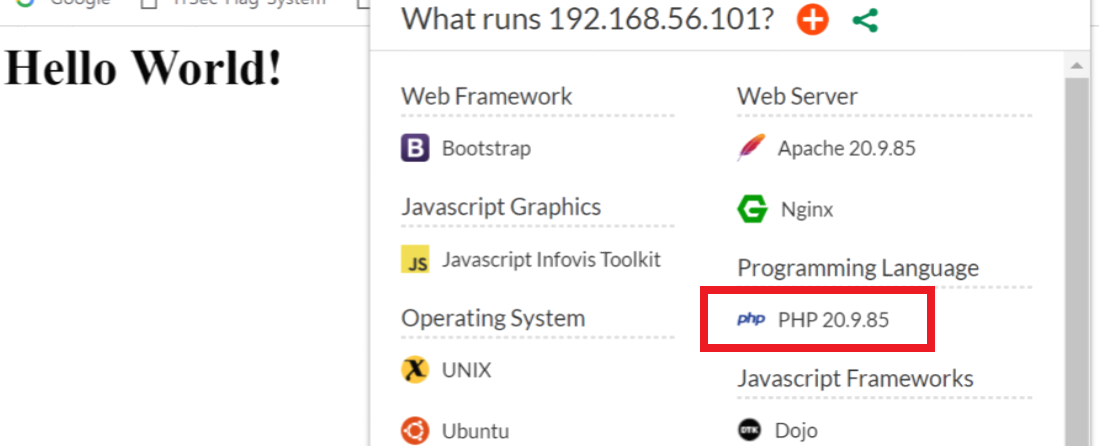}
\caption{The version scanner WhatRuns determines the version number of the PHP engine processing an HTML page.
The scanner outputs PHP 20.9.85 (red bordered value) and does not detect the additional \Patch\ information (\code{-car}), while in fact PHP 7.1.1 is running.
As a side note, version scanners are also tricked by the changed version number of Apache (also 20.9.85).}
\label{fig:whatruns}
\end{figure}

At the time of writing, PHP 7 is the highest available major software version of PHP, so the fake is quite obvious.
However, as a result every implementation of \RSI that relies on static information will yield the faked value instead the real one.
As expected, this behavior can be seen by using, for example, the version scanner WhatRuns \cite{whatruns} and visiting an example HTML page containing nothing else than 'Hello World!', see Figure~\ref{fig:whatruns}.
This demonstrates the contradiction of the current state of the art techniques and the goal they try to achieve: reliably determining the actual software version.

\section{\RSIlong}\label{sec:model}
In this section, we describe the formal framework for \RSIlong (\RSI).
All communication is done over an authenticated and encrypted channel, e.g., by employing TLS.

\subsection{System Model}
We assume that a verifier has remote access to some software hosted by a service provider and wants to determine whether the correct software is running.
Due to the fact that access is remote, the only channel of information available is to observe the input-output-behavior of the hosted software.
To this end, we model software as a finite state machine with outputs:

\begin{definition}[FSM with Outputs]
A finite state machine $\fsm$ with outputs (FSM for short) is defined by a six-tuple 
\begin{align*}
(\states,\startstate,\inputalphabet,\outputalphabet,\ftrans,\fout)
\end{align*}
where
\begin{itemize}
\item $\states$ is the set of states with $\startstate\in\states$ being the start state,
\item $\inputalphabet$ denotes the (finite) input alphabet,
\item $\outputalphabet$ denotes the (finite) output alphabet,
\item $\ftrans\colon\states\times\inputalphabet\rightarrow\states$ is the state transition function, and
\item $\fout\colon\states\times\inputalphabet\rightarrow\outputalphabet$ is the output function.
\end{itemize}
\end{definition}

At the beginning, the FSM $\fsm$ is initialized to state $\startstate$.
Then, the verifier can send input $\inp\in\inputalphabet$ to $\fsm$ and receives output $\outp\in\outputalphabet$.
More formally, assume that $\fsm$ is in some state $\state$ and that the user sends an input $\inp$.
Then, $\fsm$ produces an output $\outp=\fout(\state,\inp)$ and updates its state to $\ftrans(\state,\inp)$.

We extend the notation of the output function as follows to reflect sequences.
Given a state $\state\in\states$ and a sequence $\inputs=(\inp_1,\ldots,\inp_n)\in\inputalphabet^n$, we define
\begin{align*}
\fout(\state,\inputs)=\outputs=(\outp_1,\ldots,\outp_n)
\end{align*}
where there exists a sequence of states $\state_1,\ldots,\state_n$ such that $\state_1=\state$, $\state_{i+1}=\ftrans(\state_i,\inp_i)$ for $i=1,\ldots,n-1$, and $\outp_j=\fout(\state_j,\inp_j)$ for $j=1,\ldots,n$.
Moreover, we will usually omit the state if we refer to the start state $\startstate$, that is
\begin{align*}
\fout(\inputs):=\fout(\startstate,\inputs).
\end{align*}
Sometimes, we need to distinguish between the output functions of different FSMs.
In such a case, we will write $\foutM{\fsm}$ instead of $\fout$ to make clear that we talk about the output function of a certain FSM $\fsm$.

Recall that the only information a verifier gets about a FSM is its input-output-behavior.
Consequently, we define equivalence of FSMs according to the observable outputs. 
\begin{definition}[Equivalence of FSMs]
Consider two FSMs $\fsm$ and $\fsm'$, both defined over the same input alphabet $\inputalphabet$ and output alphabet $\outputalphabet$. 
Let $\inputs\in\inputalphabet^*$.
We say that $\fsm$ and $\fsm'$ are \emph{equivalent with respect to $\inputs$} (expressed by $\fsm\equiv_{\inputs}\fsm'$) if $\foutM{\fsm}(\inputs)=\foutM{\fsm'}(\inputs)$.
Analogously, we define for a set $\inputset\subseteq \inputalphabet^*$ that 
\begin{align*}
\fsm\equiv_{\inputset}\fsm' \quad\Longleftrightarrow\quad \fsm\equiv_{\inputs}\fsm' \forAll \inputs\in\inputset.
\end{align*}
Finally, two FSMs are equivalent, i.e., $\fsm\equiv\fsm'$, if $\fsm\equiv_{\inputalphabet^*}\fsm'$.
\end{definition}

\subsection{Attacker Model}\label{sec:attackermodel}
The audited software is operated by the prover $\prover$, i.e., software provider, according to some previously agreed on SLA with the customer or verifier.
In general, he wants to always let the latest software version.
However, this implies installing updates, software, or new configurations which results in system changes that can impact the rest of the system platform and creates a financial burden for the prover.
For example, software dependencies may change, configurations will be altered, and functionalities will be added and removed as well.
Therefore, the prover is motivated to act economically driven, that is, to reduce the number of system changes.
Hence, the rational attacker model applies to the prover.
In other words, he may alter the running software to fake a certain software version.
For example, by employing \emph{security by obscurity}, the prover might use version blinding, i.e., hide or forge the software's version.
Also, the prover could simulate some functions to convince the verifier that another software version is running.

The second party, the verifier $\verifier$, wants to identify the software operated by the prover and represents a trusted party.
When detecting misbehavior by the prover according to the SLA, i.e., a not secure software version, she might for example take legal actions.
Observe that learning which software is being run also allows to perform tailored attacks on the service.
However, we only focus on determining the identity of the software, the actions performed with this knowledge are out of scope of this work.

\subsection{\RSIlong}\label{sec:rsi}
\RSI is defined with respect to a set $\fsmset$ of pairwise non-equivalent FSMs and a target FSM $\fsmtarget$ together with two parties: a prover \prover and a verifier \verifier.
Both parties know the specifications of the FSMs contained in $\fsmset$.
The prover \prover is providing to \verifier remote access to a challenge FSM $\fsmchallenge$ and \verifier has to decide (essentially) if $\fsmchallenge=\fsmtarget$.
We are going to make this more precise now.

\RSI is divided into two phases: a setup phase and a verification phase.
We assume that prior to these phases, a set $\fsmset$ of pairwise distinct FSMs and a target FSM $\fsmtarget\in\fsmset$ have been agreed upon.

In the setup phase, \prover selects a source FSM $\fsmsource\in\fsmset$ and produces a challenge FSM $\fsmchallenge:=\produce(\fsm^*)\in\fsmset$ from some FSM $\fsm^*\in\fsmset$.
Note that $\fsmchallenge$ might not be an element of the same software family as $\fsmsource$ and $\fsm^*$ anymore, since $\produce$ might produce a FSM that is a new combination of distinct FSMs of the same software family.
In the simplest case, it holds that $\fsmchallenge=\fsmsource$, but we will also discuss the case that $\fsmchallenge$ may be a modification of $\fsmsource$, i.e., $\fsmchallenge\neq\fsmsource$. 

In the verification phase, \verifier interacts with $\fsmchallenge$ (which is run by \prover) as described above.
That is, \verifier can send inputs $\inputs$ to $\fsmchallenge$ and receives the corresponding outputs $\outputs$.
We assume furthermore that $\verifier$ can reset $\fsmchallenge$, that is to set its state back to $\startstate$.
Eventually, \verifier outputs a decision $\decision\in\{\true,\false\}$ whether it interacts with the correct FSM or not, i.e., if it holds $\fsmsource=\fsmtarget$. 

\begin{definition}[\RSI Protocol, Correctness, and Soundness]
Given the set of pairwise distinct FSMs $\fsmset$ and a target FSM $\fsmtarget\in\fsmset$, we define the \RSI protocol $\rsip$ as
\begin{align*}
\rsip\colon [\verifier\colon\inputs;\ \prover\colon \fsmchallenge, \fsmsource] \longrightarrow [\verifier\colon\decision;\ \prover\colon \bot],
\end{align*}
where the verifier \verifier sends the input sequence $\inputs$ to the FSM $\fsmchallenge$ that is controlled by the prover \prover, as is the FSM $\fsmsource$.
At the end, \verifier will output a decision $\decision\in\{\true,\false\}$ which is only true if $\fsmsource=\fsmtarget$.
The prover's output is empty ($\bot$).
We say $\verifier$ is $\rsip$-correct if it holds
\begin{align*}
\Pr{\decision=\true\mid\fsmsource=\fsmtarget}=1
\end{align*}
and \verifier is called $\rsip$-$\epsilonrsi$-sound if it holds
\begin{align*}
\Pr{\decision=\true\mid\fsmsource\neq \fsmtarget}\leq \epsilonrsi,
\end{align*}
where $\epsilonrsi$ depends on the security parameter $\secparam$.
\end{definition} 

We discuss now the different existing possibilities for enabling an \RSI scheme.
To this end, we will first focus on the case that $\fsmchallenge=\fsmsource$.
That is \verifier is given remote access to an FSM $\fsmchallenge\in\fsmset$ and has to decide based on selected inputs $\inputs$ and corresponding outputs $\outputs$ whether $\fsmchallenge=\fsmtarget$.
Obviously, the only possible approach for the verifier is to use inputs such that the outputs differ between the different FSMs in order to come to a decision.
This gives raise to the following notion.

\begin{definition}[Distinguishing Sequence]\label{def:distseq}
Let $\fsmset$ be a set of FSMs, all being defined over the same input alphabet $\inputalphabet$ and output alphabet $\outputalphabet$, and of size at least two.
We call a sequence $\inputs\in\inputalphabet^*$ a \emph{distinguishing sequence} with respect to $\fsmset$ if there exist $\fsm,\fsm'\in\fsmset$ such that $\foutM{\fsm}(\inputs)\neq \foutM{\fsm'}(\inputs)$.
Moreover, we define by $\distseq(\fsmset)\subseteq\inputalphabet^*$ the set of all distinguishing sequences.
We say that a sequence $\inputs\in\inputalphabet^*$ is \emph{non-distinguishing} if $\inputs\not\in \distseq(\fsmset)$. 
\end{definition}

The motivation behind distinguishing sequences is that they are the only means for a verifier to solve the \RSI problem, i.e., determining if $\fsmsource=\fsmtarget$.
In other words, we can restrict to verifiers that only use distinguishing sequences as inputs.

\begin{theorem}\label{thm:distseqs}
Consider an \RSI instance given by a set $\fsmset$ of pairwise distinct FSMs and a target FSM $\fsmtarget\in\fsmset$.
Let \verifier be a verifier for the \RSI problem which is $\rsip$-correct and $\rsip$-$\epsilonrsi$-sound.
Then there exists a verifier $\verifier'$ that is likewise $\rsip$-correct and $\rsip$-$\epsilonrsi$-sound and works analogously to $\verifier$ with the only exception that it omits all non-distinguishing sequences used by \verifier.
\end{theorem}

\begin{IEEEproof}
We show the claim by specifying $\verifier'$.
Recall that $\verifier'$ knows the specifications of the FSMs contained in $\fsmset$.
One consequence is that $\verifier'$ can decide for a given input sequence $\inputs$ whether it is a distinguishing sequence or not.
This algorithm uses $\verifier$ and simulates all inputs and outputs for $\verifier$ as follows.
When $\verifier$ uses as input a distinguishing sequence $\inputs\in\distseq(\fsmset)$, the algorithm $\verifier'$ forwards this input to $\fsmchallenge$ and the received output back to $\verifier$.
Whenever $\verifier$ uses a non-distinguishing sequence as input, $\verifier'$ computes the corresponding output on its own using the known specifications of the FSMs and returns the result to $\verifier$.
Eventually, when $\verifier$ outputs \true or \false, then $\verifier'$ outputs the same value.
Obviously, $\verifier'$ perfectly simulates $\rsip$ for $\verifier$ and hence ``inherits'' the same correctness and soundness properties.
\end{IEEEproof}

Theorem~\ref{thm:distseqs} shows that the task of solving the \RSI problem boils down to two procedures for the verifier:
\begin{itemize}
\item Design: Find and construct distinguishing sequences for appropriate sub-sets of $\fsmset$;
\item Strategy: Decide on a strategy on how to successively sort out candidates for $\fsmsource$ (from $\fsmchallenge$).
\end{itemize}
Note that current auditing processes for identifying the software version follow this principle.
Asking the software for its version number can be interpreted as making a distinguishing input.
However, as pointed out in Section~\ref{sec:practicalattack} it is easy for a malicious provider to tamper with the software, e.g., to fake the response to the software version request.
In the next section, we explain a novel approach for realizing \RSI even in the context of more powerful adversaries.

\section{\RFPlong}
We will first motivate \RFPlong (\RFP), explain its basic idea, then define the \RFP scheme $\rfps$ with all including protocols and procedures as well as incorporating software development specific properties, and finally give a security analysis to show that $\rfps$ solves the \RSI-problem.

Recall that according to the definition of \RSI, the provider as prover \prover grants the verifier \verifier access to $\fsmchallenge=\produce(\fsmsource)$.
That is, a provider may modify $\fsmsource$ in this process to produce $\fsmchallenge$, especially it holds then $\fsmchallenge \neq \fsmsource$.
For example, installing additional software that checks for certain inputs, e.g., that asks for the version number, and returns a forged value.
Within the framework explained in the previous section, this would mean that certain input sequences do not yield a distinguishing sequence anymore.
As we have shown, it is mandatory for any verification algorithm to design distinguishing sequences.
Thus, whether \RSI is still possible or not strongly depends on \emph{(i)} how \prover can affect the sets of distinguishing sequences and \emph{(ii)} to what extent one can still design distinguishing sequences. 

To illustrate this, let us consider the following example.
Let $\fsmset=\{\fsm_0,\fsm_1\}$ be a pair of distinct FSMs over the same input and output alphabets and let $\fsmtarget=\fsm_0$.
Assume that \prover picks as source $\fsmsource=\fsm_1$ but that $\fsm_0=\fsmchallenge=\produce(\fsm_1)$ holds.
That is, \prover applies a total conversion to $\fsm_1$ and transforms it into $\fsm_0$.
Obviously, now it becomes impossible to distinguish between the cases $\fsmsource=\fsm_0$ and $\fsmsource=\fsm_1$.
Thus, \RSI is only possible if \prover applies ``modest'' changes at most to the FSM.
This malicious actions bring us to \RFP.

\subsection{Basic Idea}\label{sec:basicidea}
We have designed \RFP for exactly such cases, where $\fsmchallenge\neq\fsmsource$.
More precisely, it is based on the observation that extensive modifications of a software would contradict the economic incentives of \prover.
For instance, the total conversion of one FSM into another explained above would not be meaningful as it would be more simple and also cheaper to install $\fsm_0$ right from the start.
The idea of \RFP is that without extensive modifications, a software keeps its \emph{intrinsic} functionalities and behavior.

The goal of \RFP is to identify the remotely running software (FSM $\fsmsource$), that is to determine its software version by communicating over basic interfaces (FSM $\fsmchallenge$).\footnote{Please note that the software provider might use \RFP to audit himself. This allows him to prove his claims and, e.g., allows for certifications.}
The result can then be compared to a target software version (FSM $\fsmtarget$) in order to determine if the prearranged software version is running.
Hence, \RFP is a building block to realize $\rsip$ and is a new approach to solve the \RSI problem.
\RFPlong incorporates the following features:
\begin{itemize}
\item \RFP performs multiple times a challenge-response protocol to successively decrease the verifier's insecurity of the audited software's version number;
\item for a challenge, \RFP uses input $\inp$ to leverage the functionality of \emph{intrinsic} processes of the audited software which are hard to simulate;
\item for a response, \RFP expects a certain output $\outp$ of the audited software within a certain time frame depending on $\inp$ and the therein included randomness $\randomness$;
\item \RFP incorporates software versioning hierarchies and software development properties to eliminate false positives;
\item at the end, \RFP evaluates all tuples of challenges and responses to determine a software version candidate set $\candidateset$;
\item in comparison to state of the art, \RFP is secure against caching, pre-compu{\-}tation, proxy-forwarding, and erroneous halts, of course in addition to determining the software version of $\fsmsource$.
\end{itemize}

%(EXAMPLES: easy: return a config value which was added in a specific version, hard: most functions that processes inputs in a certain way)
%
%RFP: 3 with depending on the randnomess of the auditor AND hard to simulate functions.
%
%Software = fsm + version label
%software family = we do not distinguish families, assume that has been happened before
%interesting: all softwares is not fsmset, fsmset is subset, is family

\subsection{Reverse Fingerprinting Scheme}
First, we will formalize the capabilities of $\prover$ for the algorithm $\produce$, that is, simulating behavior of a FSM which is different from $\fsmsource$, i.e., $\fsmchallenge\neq\fsmsource$.

\begin{definition}[Simulation-Hard Functions]\label{def:shf}
Let $\fsm\in\fsmset$ be a FSM over the input alphabet $\inputalphabet$ and output alphabet $\outputalphabet$.
We call the union of all state transition functions $\ftrans$ and output functions $\fout$ the \emph{set of functions $\funcset$} of $\fsm$ and denote this by
\[ \funcset\colon\states\times\inputalphabet\to \states\cup\outputalphabet,  \]
where $\states$ is the set of states.
By $\funcset_\fsm$ we denote the function set of the FSM $\fsm$.
Let $\fsm'=\produce(\fsm)\in\fsmset$ be a FSM with $\funcset_{\fsm'}\neq\funcset_{\fsm}$ that is also run by $\prover$.
We say that $\prover$ can perform $\produce$ if it matches his economically incentives, that is installing the real target or newer software version will cost at most the same time and/or money for $\prover$ as implementing or simulating its functions via $\produce$.
Therefore, $\funcset_\fsm'$ contains a set of \emph{simulation-hard functions} in respect to $\prover$ that can not be constructed via $\produce$, but can only be the identical functions of $\fsm$.
We denote the set of simulation-hard functions by $\funcsethard_{\fsm'} \subset \funcset_{\fsm'}$, or if the FSM is given in the context by
\[\funcsethard \subset \funcset .\]
\end{definition}

In other words, simulation-hard functions are inherent functions of a FSM $\fsm$ which can not efficiently be simulated or produced by $\prover$ via $\produce$.
That is, given a simulation-hard function $\func$ of a target FSM $\fsmtarget$, then $\fsmchallenge=\produce(\fsmsource)$ will not have the function $\func$, i.e., $\funcsethard_{\fsmtarget}\neq\funcsethard_{\fsmchallenge}$, unless $\fsmtarget=\fsmsource$.
It follows that this method allows us to distinguish FSMs.

Since \RFP aims to determine a candidate set of possible software versions of $\fsmsource$, we will write $\version_\fsm$ to denote the software version $\version$ of a FSM $\fsm$.
Let $\version_\fsm\in\outputalphabet$ such that it can be an output of $\fsm$.
For the sake of readability, we define a software $\sw$ to be the tuple of a FSM $\fsm$ and the corresponding version $\version_\fsm$, i.e. $\sw := (\fsm, \version_\fsm)$.
Let $\swall$ be the set of all softwares and let
\[ \swfamily := \swfamily_\sw = \{\sw_0,\ldots,\sw_n\} \subset \swall \]
denote the \emph{software family of $\sw$}, i.e., the set of all valid softwares and corresponding versions belonging to the same software ancestor $\sw$, e.g., all PHP versions or all WordPress versions.
We assume that the software family is always given by the environment, since the goal of \RFP is to distinguish between different software versions, but not to determine the software family.
Let further $\versions := \versions_\swfamily$ be the set of all software versions of $\swfamily$.

Recall that a distinguishing sequence $\distseq(\swfamily)$ is used by the verifier to distinguish between two FSMs of $\swfamily$ (see Definition~\ref{def:distseq}).
However, due to the attacker model of $\prover$ and the properties of $\produce$, a distinguishing sequence may not be able to distinguish between two completely different software versions.
That is, a \prover may construct $\fsmchallenge=\produce(\fsmsource)$ by simulating functions of $\fsmchallenge$ that are not given in the source FSM $\fsmsource$.
This motivates the extension of distinguishing sequences by simulation-hard functions given as follows:

\begin{definition}[Distinguishing Sequence over Simulation-Hard Functions]
Let $\swfamily$ be a given software family and $\sw\in\swfamily, \sw'\in\swall$ over the same input alphabet $\inputalphabet$ and output alphabet $\outputalphabet$.
We call a sequence $\inputs\in\inputalphabet^*$ a \emph{distinguishing series over simulation-hard functions} such that $\fout_\sw(\inputs) \neq \fout_{\sw'}(\inputs)$, where each output value $\outp_i$ depends on a simulation-hard function $\func\in\funcsethard_\sw$.
We define by $\dshard(\sw)\subseteq\inputalphabet^*$ the set of all distinguishing series over simulation-hard functions for a software $\sw$.
\end{definition}

Next, we define the \RFP scheme $\rfps$ that is initiated by the verifier $\verifier$ and consists of three procedures $\procsetup$, $\proctest$, and $\procoutput$.
An overview of $\rfps$ and its subsidiary procedures as well as protocols is depicted in Figure~\ref{fig:rfp}.
\begin{figure*}
\centering
\fbox{
\begin{tabular}{lp{20mm}l}
\multicolumn{3}{c}{\RFP Scheme $\rfps$\vspace*{0.5em}}\\
\textbf{Verifier} $\verifier$ && \textbf{Prover} $\prover$\\
\midrule
\multicolumn{3}{c}{\prulelong$\procsetup$\prulelong}\\
$\swfamily\subset\swall, \swtarget\in\swfamily, \interfaces\subseteq\interfacesall$ && $\swfamily\subset\swall, \swtarget\in\swfamily, \interfaces\subseteq\interfacesall$\\
select $\randsource$ && $\swsource\in\swfamily$\\
create $\database_\swfamily\in\databases$ && $\swchallenge \gets \produce(\swsource)\in\swall$\\
choose $\strategies$ && \\
\midrule
\multicolumn{3}{c}{\prulelong$\proctest$\prulelong}\\
$\interface\subseteq\interfaces$ && \\
$\database = \database(\swfamily, \interface)\in\database_\sw$ && \\
$\decisionlog \gets \emptyset$ && \\
$\strategy \in \strategies$ && \vspace*{0.5em}\\
\multicolumn{3}{l}{\pruleshort$\rfpp$:\pruleshort}\\
$\sw \gets \strategy(\database, \decisionlog)$ && \\
\rotatebox[origin=c]{180}{$\Lsh$} if $\sw = \emptyset$ go to $\procoutput$ && \\
$(\challenge, \expresponse, \timec) \gets\dbentry{\sw}$ && \\
$\randomness\gets_R\randsource, \timec' \gets \text{now}$ && \\
\multicolumn{3}{c}{$\pto{\challenge(\randomness)}$}\\
 && $\response\gets\swchallenge(\challenge(\randomness))$ \\
\multicolumn{3}{c}{$\pgets{\response}$}\\
$\timec'\gets\text{now}-\timec'$ && \\
\multicolumn{3}{l}{$\decision \gets \begin{cases}\true&\response=\expresponse \land \timec' \leq \timec,\\\false & \text{else}.\end{cases}$} \\
$\decisionlog \gets \decisionlog \cup (\sw, \decision)$ && \\
go to $\rfpp$ && \\
\midrule
\multicolumn{3}{c}{\prulelong$\procoutput$\prulelong}\\
$\candidateset\gets\procoutput(\swfamily, \decisionlog)$ && \\
\end{tabular}
}
\caption{Protocol representation of the \RFP scheme $\rfps$ between verifier $\verifier$ (user) and prover $\prover$ (service provider).
A database entry $\dbentry{\sw}$ relies on at least one function $\func\in\fgood(\version_\sw)\cup\fbad(\version_\sw)\cup\fugly(\version_\sw)$, where $\version_\sw$ is the software version of $\sw$ (see Section~\ref{sec:svhierarchies}).
The secret and public keys of each party are omitted, and the values of $\interfaces$, $\swtarget$, and $\swfamily$ are agreed on in the SLA beforehand.
Public values are the sets of all softwares $\swall$ and all interfaces $\interfacesall$.
Fingerprinting techniques are involved in the creation of $\database_\swfamily$.}
\label{fig:rfp}
\end{figure*}

\begin{definition}[\RFP Scheme]\label{def:rfps}
Let $\swchallenge:=(\fsmchallenge, \version_{\fsmchallenge})\in\swall$ and $\swsource:=(\fsmsource, \version_{\fsmsource})\in\swfamily$ with $\fsmchallenge = \produce(\fsmsource)$ be the softwares hosted by $\prover$.
We define the \emph{\RFP scheme $\rfps$} between $\verifier$ and $\prover$ as the consecutive procedures $\procsetup$, $\proctest$, and $\procoutput$.
At the end, $\verifier$ outputs a software candidate set $\candidateset\subset\swfamily$ with $\swsource\in\candidateset$.

We say $\verifier$ is \emph{$\rfps$-correct} if it holds
\[ \Pr{\version_\swsource \in \candidateset \mid \swchallenge=\produce(\swsource)} = 1 \]
and $\verifier$ is called \emph{$\rfps$-$\epsilonrfp$-sound} if it holds
\[ \Pr{\version_\swsource \not\in \candidateset \mid \swchallenge=\produce(\swsource)} \leq \negl(\epsilonrfp) \]
where $\negl$ is a negligible polynomial function in $\epsilonrfp$ depending the security parameter $\secparam$.
We call $\rfps$ \emph{secure} if it is $\rfps$-correct, $\rfps$-$\epsilonrfp$-sound, $\rsip$-correct, and $\rsip$-$\epsilonrsi$-sound.
\end{definition}

In the optimal case, we get $|\candidateset|=1$ in $\rfps$.
With the definition of $\rfps$, we are now able to prove that the existing four fingerprinting classes presented in Section~\ref{sec:sota} are not secure, i.e., $\rfps$-correctness or $\rfps$-soundness are not given for a verifier $\verifier$.
This can immediately be seen since none of the classes relies on inputs of $\dshard$, that is the verifier does not rely on intrinsic functions and hence can be fooled.
Please refer to Appendix~\ref{sec:ftanalysis} a detailed analysis of all state of the art fingerprinting techniques.
We will now describe the three procedures of $\procsetup$, $\proctest$, and $\procoutput$ in detail.

\subsection{The $\procsetup$ Procedure}\label{sec:rfpsetup}
This randomized procedure generates for each the verifier $\verifier$ and provider $\prover$ a public-private key pair.
If a party only deploys symmetric key schemes, the public key is simply set to $\bot$.
For the sake of brevity, we implicitly assume for each of the subsequent protocols and procedures that an involved party always uses as inputs its own secret key and the public key of the other party.
Furthermore, the set of all softwares $\swall$ is produced by the environment and the software family $\swfamily\subset\swall$ is given to both parties as well as a target software $\swtarget\in\swfamily$ (which, in practice, is chosen beforehand as part of an SLA).
By $\interface$ we denote an interface, that is a (possibly authenticated) communication channel between a party and a software.
Let $\interfacesall$ be the set of all interfaces.
We define the procedure $\procsetup$ between a verifier $\verifier$ and prover $\prover$ by
\begin{align*}
\procsetup\colon [\verifier\colon\bot;\ \prover\colon\bot] \longrightarrow [\verifier\colon\interfaces,\randsource,\databases,\strategies;\ \prover\colon\interfaces,\swsource,\swchallenge].
\end{align*}
During $\procsetup$, $\verifier$ and $\prover$ first determine which interfaces $\interfaces\subseteq\interfacesall$ they are able to use as communication channels between $\verifier$ and $\swchallenge$ and output them.
For example, $\interfaces$ may consist of FTP\cite{rfc959} and HTTP\cite{rfc2616} with according authentication and implementation details.
Different interfaces allow for different communication channels, that is, depending on the software and authorization, certain fingerprinting techniques and simulation-hard functions are available for certain interfaces only.
For example, an authenticated channel grants usually more privileges compared to an unauthenticated channel, e.g. customer and guest, respectively.
The prover $\prover$ takes the source software $\swsource\in\swfamily$ and computes the challenge software $\swchallenge = \produce(\swsource)\in\swall$, where his goal is to convince the verifier that $\swchallenge = \swtarget$.
The verifier $\verifier$ obtains a source of randomness $\randsource$, constructs a set of databases $\databases$, and defines a set of strategies $\strategies$.
An database $\database\in\databases$ in \RFP is defined as follows:

\begin{definition}[Database]\label{def:database}
Let $\swfamily$ be a given software family and $\interfaces\subseteq\interfacesall$ be some interfaces agreed on by $\verifier$ and $\prover$.
Let further $\dshard(\sw, \interface)\subseteq\inputalphabet^*_\sw(\interface)$ be the set of distinguishing sequences over simulation-hard functions for a software $\sw$ and an interface $\interface\in\interfaces$, where $\inputalphabet^*_\sw(\interface)\subseteq\inputalphabet^*_\sw$ is the restricted set of all input values for $\sw$ in respect to $\interface$.
By $\dshard(\sw, \interface)$ we denote the restriction of $\dshard(\sw)$ to an interface $\interface$.

We call $\challenge(\sw, \funcsethard_\sw, \randsource, \interface_\challenge)\in\dshard(\sw, \interface_\challenge)$ a randomized \emph{challenge} for a software $\sw\in\swfamily$ and interface $\interface_\challenge\in\interfaces$ sent from $\verifier$ to $\swchallenge$, which employs a randomness source $\randsource$ as auxiliary input and depends on at least one simulation-hard function $\func\in\funcsethard_\sw$ of $\sw$.
We will omit parameters from now on if they are clear from the context.

By $\expresponse(\challenge, \interface_\expresponse)\in\fout_\sw(\challenge)$ we denote the \emph{expected response} from $\swchallenge$ to the challenge $\challenge$ from $\verifier$ for a software $\sw\in\swfamily$.
In general, it holds $\interface_\expresponse\neq\interface_\challenge$.

Finally, we define a \emph{database} by:
\begin{align*}
\database(\swfamily, \interfaces') := \{&(\challenge, \expresponse, \timec)_\sw \mid 
\challenge = \challenge(\sw, \funcsethard_\sw, \randsource, \interface_\challenge)\in\dshard(\sw, \interface_\challenge),\\
&\expresponse = \expresponse(\challenge, \interface_\expresponse)\in\fout_\sw(\challenge), \sw\in\swfamily,\\
&\interface_\challenge,\interface_\expresponse\in\interfaces', \timec\in\N_0 \},
\end{align*}
where $\timec$ is a number of time units and $\interfaces'\subseteq\interfaces$.
A database $\database$ is called \emph{perfect}, if it holds $|\database|=|\swfamily|$.
We abbreviate by $\databases := \databases_\swfamily = \{\database(\swfamily, \interfaces')\mid\interfaces'\subseteq\interfaces\}$ the set of all databases for $\swfamily$, and by $\dbentry{\sw}:=(\challenge,\expresponse,\timec)_\sw\in\database(\swfamily, \interfaces')$ the database entry belonging to a software $\sw\in\swfamily$ for some interface.
\end{definition}

Note that since the verifier constructs $\databases$, he also computes the expected responses.
Also, two database entries for two different software versions in the same database may employ four different interfaces.
%$\dbentry{\sw}, \dbentry{\sw'}\in\database(\swfamily, \interfaces)$ and $\sw,\sw'\in\swfamily$, we can have $\interface_\challenge, \interface_\expresponse, \interface'_\challenge, \interface'_\expresponse\in\interfaces$ with $\interface_\challenge \neq \interface_\expresponse \neq \interface'_\challenge \neq \interface'_\expresponse$.
%For the sake of brevity, we write one interface for each type of messages.
A source of randomness allows to insert random values into a challenge depending on the coins of $\randsource$, which makes the challenge and especially the according correct response hard to predict.
Designing a database $\database(\swfamily,\interfaces')\in\databases$ represents the first task for $\verifier$ resulting from Theorem~\ref{thm:distseqs}.
In Section~\ref{sec:dbdesign}, we describe how to design a database for \RFP.

Finally, a strategy $\strategy\in\strategies$ describes a way how to traverse entries in a database $\database$ to successively sort out candidates in $\candidateset$ to reach $\swsource$, depending on the responses given by $\swchallenge$.
In other words, a strategy yields a distinguishing sequence over simulation-hard functions $\dshard$ for any $\sw\in\swfamily$.
We will give a formal definition of a strategy below, see Definition~\ref{def:strategy}.

\subsection{The $\proctest$ Procedure}\label{sec:rfptest}
This randomized procedure takes as input the outputs of $\procsetup$, that is the set of interfaces $\interfaces$ for $\verifier$ and $\prover$, the source of randomness $\randsource$, set of databases $\databases$, and set of strategies $\strategies$ for $\verifier$, and the challenge software $\swchallenge$ for $\prover$.
At the end, the verifier outputs set of decisions $\decisionlog$ for each software version that has been tested against $\swchallenge$.
We now formally define testing, that is on processing an entry of a database, the verifier performs a challenge-response protocol $\rfpp$ which we define as follows:

\begin{definition}[\RFP Protocol]\label{def:rfpp}
Let $\swfamily$ be a given software family, $\decision\in\{\true,\false\}$ be a decision set, and $\timec'$ be a measure of time units.
We define the \emph{\RFP protocol} between the verifier $\verifier$ and prover $\prover$ as
\begin{align*}
\rfpp\colon [\verifier\colon\dbentry{\sw},\randsource;\ \prover\colon\swchallenge] \longrightarrow [\verifier\colon\decision;\ \prover\colon\bot],
\end{align*}
for a database entry $\dbentry{\sw}\in\database(\swfamily,\interfaces)$ with $\sw\in\swfamily$ and a set of interfaces $\interfaces$, and a randomness source $\randsource$.
$\rfpp$ consists of three steps for the verifier $\verifier$, which we call an \emph{\RFP test}:
\begin{enumerate}
\item send a randomized challenge $\challenge(\randomness)\in\dbentry{\sw}$ with $\randomness\in_R\randsource$;
\item receive a response $\response$ from $\swchallenge$ after $\timec'$ time units, otherwise set $\response=\bot$ if $\timec'$ passes some time threshold;
\item fetch $\expresponse\in\dbentry{\sw}$ and output $\decision$, where $\decision\gets\true$ iff $\response=\expresponse$ and $\timec'\leq\timec$, otherwise $\decision\gets\false$.
\end{enumerate}
\end{definition}

We can now perform a single \RFP test via $\rfpp$, which yields a single outcome $\decision$ for a certain software $\sw$.
First, observe that a challenge $\challenge$ of a database entry depends on $\sw$, which in fact means that the intrinsic behavior of $\sw$ is being tested.
However, as we have shown in Section~\ref{sec:sota}, state of the art fingerprinting techniques will not be sufficient.
Hence, we introduce new fingerprinting techniques fulfilling \RFP security, see Section~\ref{sec:newft}.
Second, observe that $\rfpp$ will not yield the software version of $\swsource$.
That is, usually, the amount of software features grows over time with the software versions of $\swfamily$, hence testing a single software $\swfamily$ in $\rfpp$ will indeed yield true or false, i.e., if a certain simulation-hard function can be performed correctly, but is not sufficient to determine the software version of $\swsource$.
Consequently, if it holds $\decision=\false$, we learn that the software version is smaller than the one of $\sw$, and in the case of $\decision=\true$, $\swsource$ might be equal to $\sw$, but there might also be a newer version $\sw'\in\swfamily$ with $\version_{\sw'} > \version_\sw$, for which $\rfpp$ also yields $\decision=\true$.
Hence, as a core part of \RFP, $\rfpp$ is performed multiple times.

\RFP leverages a database of challenges and expected responses in combination with a strategy to produce a certain distinguishing sequence of simulation-hard functions $\dshard$ for each $\sw\in\swfamily$.
In other words, each $\sw\in\swfamily$ can be determined by testing the simulation-hard functions of multiple, different software versions in $\swfamily$ until we get a lower and upper limit for the software version of $\swsource$.
When performing multiple tests using $\rfpp$, the verifier tracks each software-decision-pair $(\sw, \decision)$ in a \emph{decision log} $\decisionlog$.
We combine the multiple tests by defining the procedure $\proctest$ as follows:
\begin{align*}
\proctest\colon [\verifier\colon\interfaces,\randsource,\databases,\strategies;\ \prover\colon\interfaces,\swchallenge] \longrightarrow [\verifier\colon\decisionlog;\ \prover\colon\bot].
\end{align*}
Essentially, $\proctest$ consists of multiple executions of $\rfpp$, where the order of the softwares that are being tested by $\rfpp$ is defined by a strategy $\strategy\in\strategies$.
We will now explain the general concept of a strategy in detail.

At the beginning, when performing $\rfpp$ for the first time, $\verifier$ needs to choose a $\sw_0\in\swfamily$ to start from.
Then, depending on the outcome $\decision_0$ of $\rfpp$ for $\sw_0$ and a given database, the task of choosing the next software $\sw_1$ is to be solved.
In other words, $\verifier$ searches for $\sw_1$ depending on $(\sw_0, \decision_0)$.
In general, we want to determine the next software that has to be tested to reduce the set of possible software candidates for the source software $\swsource$.
More formally, we are looking for $\sw_i$ for a given log $\decisionlog = \{(\sw_0, \decision_0), \ldots, (\sw_{i-1}, \decision_{i-1})\}$.
We solve this problem by using a strategy which we formally define as follows:

\begin{definition}[Strategy]\label{def:strategy}
Let $\swchallenge$ be the challenge software hosted by $\prover$ and $\decisionlog = \emptyset$.
We denote by $\decision_{\rfpp(\sw)}$ the output of $\rfpp$ for a software $\sw\in\swfamily$, a given randomness source $\randomness$, a database $\database(\swfamily,\interfaces)\in\databases$ for a software family $\swfamily$ and interfaces $\interfaces$.

Then, a \emph{strategy} is defined as a method that describes, based on $\decisionlog$ and $\database(\swfamily,\interfaces)$, which software $\sw\in\swfamily$ is selected for the next run of $\rfpp$.

Let $\candidateset,\candidateset'\subset\swfamily$ be candidate sets of possible software versions containing $\swsource$, deduced from the current decision log $\decisionlog$ and the upcoming one $\decisionlog\cup(\sw, \decision_{\rfpp(\sw)})$, respectively.
We call a strategy \emph{efficient} if it holds $|\candidateset'| < |\candidateset|$.

If there exists no further $\sw\in\swfamily$ such that the number of elements in $\candidateset$ can be reduced (or a certain threshold of executions of $\rfpp$ has been performed), the strategy commands to end the procedure $\proctest$.
\end{definition}

Please observe that the dependency on a database for a strategy is required, since the database may not be perfect, i.e., entries for some softwares may be missing.
While selecting \emph{any} software version for testing is easy, finding \emph{efficient} strategies is desired since this reduces the time required as well as communication and storage overhead for $\verifier$.
Hence, we say an \emph{optimal} strategy is represented by requiring the minimum amount of tests to find $\swsource$, i.e., the distinguishing series of simulation-hard functions $\dshard$ for the optimal strategy is at most as long as any other one $\dshard'$.
Note that if $\decisionlog=\bot$, choosing the starting software version $\sw_0$ is crucial and influences the $\dshard$ strongly.
For example, starting with the highest software version in $\database$ and successively testing step by step smaller software versions will only be an efficient strategy if $\swsource$ has a very recent software version.
At the end, $\proctest$ employs multiple executions of $\rfpp$ which produces over time a distinguishing sequence over simulation-hard functions of $\swsource$, i.e., $\dshard(\swsource)$.
Creating a strategy $\strategy\in\strategies$ represents the second task for $\verifier$ resulting from Theorem~\ref{thm:distseqs}.
We give more information on how to construct strategies in Section~\ref{sec:strategies}.

Observe that given two softwares $\sw,\sw'\in\swfamily$ with $\version_\sw < \version_{\sw'}$, this does not necessarily imply that $\sw$ is older or more insecure than $\sw'$ and furthermore it might hold $|\funcset(\sw)| > |\funcset(\sw')|$.
This is important for the design of a strategy, since there is often more than one possible ``next software version'' to choose from, i.e., the $\dshard$ is not always continuous with respect to the software version number.
We cope with the problem of software version hierarchies and software development in Section~\ref{sec:svhierarchies}.

\subsection{The $\procoutput$ Procedure}
This procedure gathers the results that have been produced by $\proctest$ and outputs a candidate set $\candidateset\subset\swfamily$ which contains $\swsource$.
Since we have stored the decision outputs of all $\rfpp$ runs in $\decisionlog$, this procedure consists of two steps:
first, a lower and upper bound of software versions is computed based on $\decisionlog$ and the software version hierarchy.
Second, given these bounds, we determine all $\sw\in\swfamily$ that lie in between the bounds, finally resulting in $\candidateset$.
We denote both steps as the procedure $\procoutput$ defined as follows:
\begin{align*}
\procoutput\colon [\verifier\colon\decisionlog;\ \prover\colon\bot] \longrightarrow [\verifier\colon\candidateset;\ \prover\colon\bot].
\end{align*}
This procedure is performed by $\verifier$ and independent of $\prover$ with the input being the log of decisions $\decisionlog$ coming from the procedure $\proctest$.
With $\procoutput$, the verifier $\verifier$ outputs a candidate set $\candidateset\subset\swfamily$ with $\swsource\in\candidateset$, solving the \RSI-problem.
Note that the cardinality of $\candidateset$ may be greater than one, since there can be $\sw,\sw'\in\swsource$ with $\sw\equiv\sw'$, as mentioned above in Section~\ref{sec:rfpsetup}.

\subsection{Distinguishung Software Hierarchies}\label{sec:svhierarchies}
Recall that the goal of the verifier is to leverage the set of simulation-hard functions of each software $\sw\in\swfamily$ to construct a distinguish series, i.e., to distinguish between single software versions.
However, as mentioned before, there exists no simple ordering of versions of the softwares of a software family.
In other words, how does the verifier know which software version is based on other software versions?
Here, two challenges arise due to software development properties, namely software branches and deprecated functions.
Together with simulation-hard functions, these three describe a \emph{software hierarchy} over $\swfamily$.
Usually, the set of functions of a newer software version contains all functions of older software versions.
However, due to software hierarchies, the version number of one software may be smaller compared to another, while it might be a newer software.
More formally, while there is a ordering on version numbers, this does not translate to the set of functions of a software, i.e., there exists $\sw,\sw'\in\swfamily$ with $\version_\sw>\version_{\sw'} \not\Rightarrow \funcset_\sw\supset\funcset_{\sw'}$.
Plus, when a strategy $\strategy\in\strategies$ chooses the next software $\sw\in\swfamily$ to test with $\rfpp$, then this decision is based on a database $\database\in\databases$, which entries must reflect the software hierarchies.
Therefore, this raises the question on how to map software hierarchies to version fingerprinting, since higher software versions does not always implicates more or newer functions.
We will now describe how to distinguish between software versions in the presence of software hierarchy including software development properties.

Since we will now focus on the version properties of a software, we are using our existing notation but refer to a software via its version $\version_\sw$ instead the software $\sw$ itself.
Let $\funcset(\version_\sw):=\funcset_\sw$ for a software $\sw\in\swfamily$.
In software development, usually the following holds:
After implementing a functionality $\func\in\funcset_\sw$ in a software $\sw$, usually all following versions of the same software family will also have implemented $\func$.
More formally,
\begin{align}
\func\in\funcset(\version)
\Rightarrow \func\in\left(\funcset(\version)\cap\funcset(\version')\right) \forAll \version' \geq \version,
\label{eq:f}
\end{align}
where $\version = \version_\sw$ and $\version'\in\versions$.

Recall that in the protocol $\rfpp$, the verifier $\verifier$ sends a randomized challenge $\challenge$, receives the response $\response$ by $\swchallenge$, and measures the time passed in between as well as compares $\response$ to the expected response $\expresponse$.
Since $\verifier$ is essentially testing if $\swchallenge$ can successfully perform a certain, inherent simulation-hard function of a software $\sw$, it follows due to equation~\eqref{eq:f} that if $\decision=\true$ we have $\version_\swchallenge \geq \version_\sw$, i.e., the version of the software provided ($\swchallenge$) is at least the on of the software compared to ($\sw$).
We denote by $\decision_{\rfpp(\sw, \func)}$ the output of $\rfpp$ depending on software $\sw\in\swfamily$ and an employed challenge derived from a function $\func\in\funcset_\sw$:
%Hence, we can write the output $\decision_{\rfpp(\sw, \func)}$ as
\begin{align*}
\decision_{\rfpp(\sw, \func)} \gets \begin{cases}
\true & \text{if }\version_\swchallenge \geq \version_\sw,\\
\false & \text{if }\version_\swchallenge < \version_\sw.
\end{cases}
\end{align*}
See Remark~\ref{rem:decision} on why we decided to use this specific definition of $\decision$.
%For brevity, we will denote by $\viewvtest{\sw}$ the view on the output $\decision$ of $\rfpp$ by $\verifier$ and a software $\sw\in\swfamily$.
%Further, let $\challenge(f)$ be the challenge that depends on the function $\func\in\funcset_\sw$.

At the beginning, i.e., when no \RFP test has been performed yet, the candidate set of all possible software versions $\candidateset$ will be equal to all versions of $\sw$, i.e., $\candidateset=\swfamily$.
Employing $\rfpp$ multiple times to test for different softwares $\sw\in\swfamily$ using a database will yield a reduced candidate set $\candidateset'\subset\candidateset$, since the $\verifier$ learns if $\swchallenge$ supports a certain, version intrinsic function of $\sw$, or not.
The final candidate set will be as small as possible, in the best case it contains only one element.

\begin{definition}[Simulation-Hard Intrinsic Functions]\label{def:ishf}
For $\sw\in\swfamily$, let
\begin{align}
\funcsetgood(\sw) := \left\{ \funcsethard_\sw\setminus\bigcup_{\substack{\sw'\in\swfamily\\\version_{\sw'} < \version_\sw}}\funcset_{\sw'} \right\}. \label{eq:1}
\end{align}
be the set of all \emph{simulation-hard intrinsic} functions of $\sw$ that are new for this software version, i.e., did not exist in a version before.
We set $\funcsetgood(\version_\sw) := \funcsetgood(\sw)$.
\end{definition}

Observe that all next released software versions, i.e., with versions greater than $\version_\sw$, also incorporate $\funcsetgood(\sw)$, i.e., $\funcsetgood(\version) \subset \funcsetgood(\version')$ with $\version < \version'$, $\version,\version'\in\versions$.

Besides checking for a certain simulation-hard intrinsic function $\func\in\funcsetgood(\sw)$ that is characteristic for that software version (and following), there are two properties of software versioning and development to take into account:
\begin{enumerate}
\item Software branches, i.e., same function changes are implemented for different version numbers, and
\item Deprecated and removed functions, i.e., functions that exist only for a few versions of $\swfamily$.
\end{enumerate}
For each of both properties, equation~\eqref{eq:f} does not hold anymore.
Hence, we will now give new ways to differentiate between all three cases to determine the outcome of $\rfpp$.

For the following three subsections, we set $\decision_{\rfpp(\version_\sw, \func)} := \decision_{\rfpp(\sw, \func)}$, where the challenge employed in $\rfpp$ depends on the function $\func\in\funcsetgood(\sw)$.
Further, let $\sw\in\swfamily$ be a given software and let $\version=\version_\sw$.

\subsubsection{Intrinsic Functions}
For a $\func\in\funcsetgood(\version)$ let the challenge $\challenge$ in $\rfpp$ depend on $\func$.
Then, $\func$ is an intrinsic simulation-hard function introduced in version $\version$ of $\swfamily$, see also Definition~\ref{def:ishf}.
We define the \emph{intrinsic set of $\sw$} as the set with the intrinsic properties, that is
\begin{align}
\fgood(\version) := \funcsetgood(\version).\label{eq:good}
\end{align}
Then, with $\func\in\fgood(\version)$, we get 
\begin{align*}
\decision_{\rfpp(\version, \func)} =
\begin{cases}
\true & \text{if }\version_\swchallenge \geq \version, \\
\false & \text{if }\version_\swchallenge < \version.
\end{cases}
\end{align*}
We now come to the other remaining cases.

\subsubsection{Branch Handling}
For $\version>\version'$ with $\version'\in\versions$, we might have $\funcsetgood(\version) = \funcsetgood(\version')$ because both versions are maintained at the same time, but live on different branches, e.g., same value for \Major but different values for \Minor or \Patch.
So given a function $\func\in\fgood(\version)$ as in the case for intrinsic functions, we now at the same time also have $\func\in\fgood(\version')$, which is a contradiction to Definition~\ref{def:ishf}.
Hence, we need to cope in a new way with this case which we dub \emph{branch detection} for versions of $\swfamily$.

To solve this problem, we additionally store for the entry $\dbentry{\sw}$ in the database a set of branch detecting tests.
These are referrers to other database entries, essentially to a version where the branch can be determined using $\funcsetgood$.
We now have the following relation between two versions:
\begin{align}
\funcsetgood(\version) = \funcsetgood(\version'), \quad \version>\version'.\label{eq:2}
\end{align}
Recall that $\funcsetgood(\version)$ are all simulation-hard intrinsic functions that have been added in version $\version$ to $\swfamily$.
Hence, equation~\eqref{eq:2} states that the same features were added to both software versions $\version$ and $\version'$.
First, let the database entries for $\version$ and $\version'$ contain a challenge depending on $\func\in\funcsetgood(\version)$.
Second, let the database entry for $\version$ contain a branch detecting test referring to version $\widehat{\version}$.
Depending on the difference between $\version$ and $\version'$, $\widehat{\version}$ has the same values \Major and \Minor as $\version$, but \Patch is set to zero; or same \Major as $\version$ and both \Minor and \Patch are set to zero.
Therefore, $\widehat{\version}\neq\widehat{\version'}$ means that the softwares of $\version$ and $\version'$ refer to different branch detecting test versions as their ancestors.
To illustrate this, example values can be as follows: $\version = (7,2,9)$, $\version' = (7,1,21)$, and $\widehat{\version} = (7,2,0)$.
We define the \emph{branched set of $\sw$ in respect to $\version'$} as
\begin{align}
\fbad(\version, \version') := \left\{f\in\funcsetgood(\version) \cap \funcsetgood(\version') \mid \version>\version' , \widehat{\version}\neq\widehat{\version'}\right\}.\label{eq:bad}
\end{align}
Then, with $\func\in\fbad(\version, \version')$, $g\in\fgood(\widehat{\version})$, we get
\begin{align*}
\decision_{\rfpp(\version, \func)} =
&\begin{cases}
\true & \text{if }\version_\swchallenge \geq \version, \\
\true & \text{if }\version_\swchallenge \geq \version', \\
\false & \text{if }\version_\swchallenge < \version'.
\end{cases}
\end{align*}
In case of a \true result, we perform $\rfpp$ a second time, but with parameters $\widehat{\version}$ and $g$, and get
\begin{align*}
\decision_{\rfpp(\widehat{\version}, g)} =
&\begin{cases}
\true & \text{if }\version_\swchallenge \geq \version, \\
\false & \text{if }\version_\swchallenge \geq \version'.
\end{cases}
\end{align*}
Please note that if there is no $g\in\fgood(\widehat{\version})$, we recursively test $g\in\fbad(\widehat{\version}, \widehat{\version}')$ or $g\in\fugly(\widehat{\version}, \widehat{\version}')$ (see below) for according $\widehat{\version}'$ until we reach a version $g\in\fgood(\widehat{\version})$.
Further note that if the branch detection test versions are equal of both versions, i.e., $\widehat{\version}\neq\widehat{\version'}$, then both versions are indistinguishable, i.e., $\sw\equiv\sw'$ with $\version=\version_\sw$ and $\version'=\version_{\sw'}$.

\subsubsection{Deprecated Functions}
Suppose there exists an $\func\in\fgood(\version)$, but we also have $\func\not\in\fgood(\version')$ with $\version'>\version$.
This is a contradiction to the definition of $\fgood$.
Futhermore, $\func$ also cannot be an element of $\fbad$.
In other words, $\func$ is only implemented in $\version$.
Therefore, we are in the third case representing \emph{deprecated simulation-hard} functions.
These are functions which only exist for a certain amount of consecutive versions, being removed from $\swfamily$ afterwards.

In other words, $\func$ exists in a version $\version$ for all versions up to another version $\version''$, but not at any higher version $\version'$.
That is, we have $\func\in\funcsethard(\version)$, $\func\in\funcsethard(\version'')$, and $\func\not\in\funcsethard(\version')$ with $\version \leq \version'' < \version'$.
We define the \emph{depcrecated set of $\sw$ with respect to $\version'$} as
\begin{align}
\fugly(\version, \version') := \funcsetgood(\version)\setminus\funcsetgood(\version').\label{eq:ugly}
\end{align}
Note that a function $\func\in\fugly(\version, \version')$ can be re-introduced after version $\version'$ later on, which requires to analogously adapt this equation.

To incorporate this case in $\rfpp$, we add to the database for entry $\dbentry{\sw}$ a referrer to a function-deprecated test version, representing a test for version $\version'$, being the version of the first released software $\sw'$ where $\func$ is not anymore part of $\swfamily$ anymore afterwards.
Then, with $\func\in\fugly(\version, \version')$, $g\in\fgood(\version')$, we get
\begin{align*}
\decision_{\rfpp(\version, \func)} =
&\begin{cases}
\true & \text{if }\version \leq \version_\swchallenge < \version', \\
\false & \text{if }\version_\swchallenge < \version, \\
\false & \text{if }\version_\swchallenge \geq \version',
\end{cases}
\end{align*}
In case of a \false result, we perform $\rfpp$ a second time, but with the parameters $\version'$ and $g$, which yields
\begin{align*}
\decision_{\rfpp(\version', g)} =
&\begin{cases}
\true & \text{if }\version_\swchallenge \geq \version', \\
\false & \text{if }\version_\swchallenge < \version.
\end{cases}\\
\end{align*}
This may, as in the case for branch detection, be done recursively.
Note that there may exist $\func\in(\fbad(\version, \version')\cap\fugly(\version, \version'))$ for some $\version'\in\versions$, representing functions that have been added to different branches and are deprecated at the same time.

\subsection{Security Analysis}
For the security analysis of the \RFP scheme $\rfps$ our goal is to show that it is secure as described in Definition~\ref{def:rfps}.
In other words, a verifier $\verifier$ of $\rfps$ must fulfill correctness and soundness in respect to both $\rsip$ and $\rfps$, respectively.

\begin{IEEEproof}
We first show that $\rfps$ is a building block for $\rsip$.
Assume that $\verifier$ is correct and sound according to $\rfps$.
In the SLA, both parties agreed on the target software $\swtarget\in\swfamily$.
At the end of $\rfps$, $\verifier$ outputs a candidate set $\candidateset\subset\swfamily$ with $\swsource\in\candidateset$.
Then, $\verifier$ can compare $\swtarget$ to $\candidateset$ and outputs $\true$, if it is contained therein, otherwise $\false$.
Hence, $\verifier$ is correct and sound according to $\rsip$.

Now we are left to show that $\rfps$-correctness and $\rfps$-$\epsilonrfp$-soundness are given for $\verifier$.
Since $\rfps$ executes tests via $\rfpp$ successively for different softwares $\sw\in\swfamily$, after some time (depending on the strategy) a certain amount of softwares have been tested such that $\proctest$ yields a decision log $\decisionlog$ which, input into $\procoutput$, results by computing upper and lower bounds of software versions in a final candidate set $\candidateset$.
This is due to the upper and lower bounds reducing the cardinality of $\candidateset$ for each step of an efficient strategy, enclosing only $\swsource$ at the end.
This holds as long as the strategy reduces the cardinality of $\candidateset$ at any time, it must not be efficient, but it must be possible to eventually test any software version.
Next, let us consider $\rfps$ leveraging Class~4 fingerprinting techniques, i.e., security flaw-depending techniques (see Section~\ref{sec:sota}).
This may call a transition $\ftrans$ or output function $\fout$ of $\swchallenge$ such that the software runs into an error state $\bot$ and abruptly halts before $\rfps$ ends.
To prevent this, $\rfps$ only employs fingerprinting techniques without transitioning $\rfps$ into $\bot$, see Section~\ref{sec:newft}.
Therefore, $\rfps$-correctness is given due to successive tests of inherent functions of $\swtarget$ on $\swchallenge$ based on named fingerprinting techniques producing a successively reduced candidate set $\candidateset$.

For $\rfps$-soundness, we need to model the capabilities of $\prover$ to act malicious.
Essentially, there are the following three ways:

\subsubsection{Simulating Functions}
Recall that $\verifier$ communicates with $\swchallenge$, which the prover $\prover$ uses to cover direct access to $\swsource$ and to mimic $\swtarget$.
To do so, he sets $\swchallenge=\produce(\swsource)$ and may simulate functions that are not part of $\swsource$ (but of $\swtarget$).
However, following from the Definitions~\ref{def:shf} and \ref{def:ishf}, altering $\swsource$ is only possible for functions that are \emph{not} simulation-hard.
But altering or simulating such functions $\func\not\in\funcsetgood$ will not affect $\rfpp$, since it only tests for functions $\func\in\fgood\cup\fbad\cup\fugly$ which are all simulation-hard.
Hence, a malicious prover $\prover$ can not fool the verifier $\verifier$ by simulating functions.

\subsubsection{Caching or Pre-Computation of Responses}
After $\verifier$ did audit $\prover$ for a software $\swtarget$ once, the $\prover$ can fool the verifier for the next audit as follows:
He records all challenges and responses and stores them in a database.
Next, he sets $\swsource$ and $\swchallenge$ to any softwares he wants.
Then, if $\verifier$ audits $\prover$ again, e.g., in form of a regular scheduled auditing process, $\prover$ just replies with the entries from his database.
Recall that a challenge $\challenge$ in $\rfpp$ depends on the randomness source $\randsource$ that is either chosen by $\verifier$.
That is, two challenges for the same software version $\sw\in\swfamily$, same interface $\interface_\challenge\in\interfaces$, and same simulation-hard function $\func$ will always be different, i.e.,
$\challenge(\sw, \randomness, \interface_\challenge) \neq \challenge(\sw, \randomness', \interface_\challenge),$
where $\randomness,\randomness'\in_R\randsource$ are different with overwhelming probability.
Hence, a malicious prover $\prover$ can not fool the verifier $\verifier$ using caching or pre-computation of responses.

\subsubsection{Proxy-Forwarding of Challenges}
Assume that in addition to $\prover$ there exists another prover $\prover'$ who has installed $\swtarget$, that is the software in the version agreed on in the SLA.
If $\verifier$ sends a challenge $\challenge$ to $\prover$, he forwards $\challenge$ to $\prover'$, who in turn computes the correct response $\response$.
Latter is then sent over $\prover$ to $\verifier$, who would then come to the conclusion, that $\prover$ indeed runs the correct software version.
However, forwarding $\challenge$ and $\response$ requires a certain amount of time $\timec'$.
Since there is an upper time limit $\timec\in\dbentry{\sw}$ for each challenge (see Definition~\ref{def:database}), this value can be set conservatively, such that forwarding is not possible anymore without being detected, i.e., taking too much time such that $\rfpp$ will output $\decision=\false$.
Hence, a malicious prover $\prover$ can not fool the verifier $\verifier$ by forwarding challenges.

Since all malicious options for the prover $\prover$ are not successful, the verifier $\verifier$ fulfills soundness in respect to $\rfps$.
In conclusion, $\verifier$ is correct and sound with respect to both $\rsip$ and $\rfps$ and therefore $\rfps$ is secure and also solves the \RSI-problem.
\end{IEEEproof}

\section{Design \& Implementation of \RFP}
In this section, we introduce three new fingerprinting techniques for $\rfps$ to design challenges and expected responses.
We furthermore delve into the design of a database and strategies, giving examples each.
Finally, we show the advantages of the \RFP structure for implementation and employ our own prototype to remotely audit software.

\subsection{Secure Fingerprinting Techniques}\label{sec:newft}
As we have shown in Section~\ref{sec:sota}, current state of the art fingerprinting techniques are not secure, that is they can be fooled by a malicious provider.
In Appendix~\ref{sec:ftanalysis} we give a detailed security analysis for each class of the fingerprinting techniques.
We have used these techniques directly as part of $\rsip$, but note that they still remain insecure when used as part of $\rfpp$.
This is due to missing features such as time constraints, randomization of challenge-response-pairs, and relying on simulation-hard functions.
Consequently, none of these should be used to audit software.

We will now introduce three new fingerprinting techniques which all, implemented in $\rfpp$, lead to a secure scheme $\rfps$.
Furthermore, these techniques belong to a new class which we call \emph{\ftclassfive}.
Note that a fingerprinting technique can directly be seen as a design instruction for challenges and according expected responses.

\subsubsection{\ftrequestresponse}\label{sec:requestresponse}
The verifier requests the software $\swchallenge$ to perform a certain simulation-hard function $\func$ in a specific way with some input parameters $\phi$.
This means that $\func$ can be nested, e.g., iterated, or that it is used as part of a larger function or control structure.
At least one of the three components must be randomized and the response by $\swchallenge$ must depend on this randomness.
That is, the output space of $\outputalphabet^*$ of $\swchallenge$ must be polynomial in the size of the randomness space of the randomness source $\randsource$ in respect to $\func$.

As an example, let \code{diffDate(a,b)} be a simulation-hard function that returns the difference between two date objects \code{a} and \code{b}, where a date object is defined as in RFC 3339\cite{rfc3339}.
Then, the challenge sends two random date objects \code{a} and \code{b} as challenge and checks if the response reflects the correct implementation of the function \code{diffDate}.

\subsubsection{\fterrorhandling}\label{sec:errorhandling}
In contrast to \fterrorcode, we are not interested in the error itself, but rather \emph{when} the error occurs and \emph{why}, i.e., which circumstances lead to cause the error.
Here, the sent challenge depends on a simulation-hard function with parameters which produces an error for a given software version of the software family.
Randomization is done as in \ftrequestresponse, and the response consists of the erroneous behavior of the called function.
Note that \fterrorhandling can be realized by using \ftrequestresponse in a certain way, hence they are in the same fingerprinting technique class.

As an example, let \code{diffDate} be a simulation-hard function as above but that does \emph{not} support the input in RFC~3339 format for a $\sw\in\swfamily$.
Then, given a randomized input in RFP~3339 format, the function will throw a certain error depending on the input value which allows deducing the software version.

\subsubsection{\ftsecurityexploitnd}\label{sec:securityexploitnd}
This technique exploits non-destructive security vulnerabilities in a software with randomized function parameters.
If no update has been applied, the exploit will be successful and hence leak information about the installed software version.
This deviates in two points compared to \ftsecurityexploit:
First, we only use security vulnerabilities that are non-destructive, that is they do not damage the system and especially do not crash the system leading to an abrupt halt.
Second, this allows for testing multiple non-destructive security flaws in order to retrieve information about the version number.
Note that this fingerprinting technique can be realized by using \ftrequestresponse, hence they are in the same class of fingerprinting techniques.

As an example, an authentication check may not be performed correctly, i.e., with a security flaw, hence allowing access to otherwise protected information.

\subsection{Database Design}\label{sec:dbdesign}
A \RFP database $\database(\swfamily, \interfaces)$ is defined over the software family $\swfamily$ and a set of interfaces $\interfaces$.
It consists of triples $(\challenge, \expresponse, \timec)$ with challenge $\challenge$, expected response $\expresponse$, and time constraint $\timec$, see Definition~\ref{def:database} for details.
As stated in Section~\ref{sec:rfptest}, designing such a database is one of the main tasks of the verifier.
The design comprises four steps which we will now explain in detail.

\subsubsection{Finding Simulation-Hard Functions}
The first goal is to find at least one simulation-hard function $\func\in\funcsetgood(\sw)$ for each $\sw\in\swfamily$ and a given set of interfaces $\interfaces\subseteq\interfacesall$.
Usually, the interfaces are known a priori to designing the database, that is how a user connects to the software or service.
Therefore, we will focus on finding suitable functions.

Following from Section~\ref{sec:svhierarchies}, determining software versions comes down to requiring intrinsic functions to distinguish between them, i.e., by relying on tests via $\rfpp$ employing functions out of $\fgood$, $\fbad$, and $\fugly$.
Note that $\fbad$ and $\fugly$ will eventually also require intrinsic functions of $\fgood$.
Hence, we will need simulation-hard intrinsic functions for each software version as basis for a \RFP database, which also need to be able to be randomized in some way, e.g., by parametrization or structure.
To find these functions, a good place to start is to research the version history or release notes of $\swfamily$, that is the list of individual changes from one version to the next or previous one.
They are often provided by the software developers, e.g., PHP changelog\cite{phpchangelog}, but can also be extracted from repositories such as GitHub\cite{github}.
Additionally, compatibility changes also help finding intrinsic functions, e.g., Firefox site compatibility\cite{firefoxcompatibility}.
If $\swfamily$ is closed source, i.e., the current software code and/or version history are not available, the verifier can try to build a version history by himself by analyzing other sources such as news and developer blogs for $\swfamily$.
At the end, the verifier might stick to the fingerprinting technique \ftsecurityexploitnd based on known, e.g., leaked, security exploits.

By comparing the properties, functions, and behavior of distinct (especially consecutive) releases of the software versions, the verifier selects those functions which are only available up until a certain software version and are no longer available (deprecated functions), or which are beginning to be part of the software starting at a certain software version (new intrinsic functions).
In addition to removed or new functions, one may also choose altered functions which then also posses new intrinsic properties.
Recall that different interfaces allow for different functions being accessed, increasing or decreasing possibly the set of available functions to gather for a database.
At the end, we have a set of simulation-hard functions $\fgood(\sw)$ for each $\sw\in\swfamily$, or, if due to software hierarchies there is no such function for $\sw$, the sets $\fbad(\sw)$ and $\fgood(\sw)$ will complement the respective entry in the database.
If, however, (i) there is no function available for a certain $\sw$, a test later on via $\rfpp$ will not return \false, since the provider did not fail at the test, since for $\rfps$ it holds $\sw\equiv\sw'$ for some other software version $\sw'$.
And (ii) we can not determine if $\sw$ is \emph{not} running, i.e., we set $\decision_{\rfpp(\sw)}=\true$ if $\fgood(\sw)=\emptyset$ (and $\fbad(\sw), \fugly(\sw)$ are also empty).

\subsubsection{Computing Challenges and Expected Responses}
Based on the functions of the set $\fgood(\sw)$, which have been gathered in the previous step, we compute the triple $(\challenge, \expresponse, \timec)_\sw$ for each $\sw\in\swfamily$.
As mentioned in Section~\ref{sec:rfpsetup}, a challenge also depends on a randomness source $\randsource$.
The verifier selects $\randsource$ to be a cryptographic secure random number generator, that is it is sufficiently difficult do guess the next output based on previous outputs.
He includes $\randsource$ in the design of the challenge $\challenge$ to randomize the parameters or structure of $\func$, or both, later on.
The challenge itself is stored in the database with placeholders for the randomness $\randomness\in\randsource$.
They are being replaced dynamically as part of $\rfpp$ just before sending the challenge to $\prover$.
Obviously, the details depend heavily on $\func$ itself.
In general, the goal is to let the provider do some randomized work over $\func$ which can only be solved in the given time frame if $\func$ is implemented and working as intended.

Next, the verifier simulates a perfect software provider which runs a software with version $\sw$.
This allows him to compute the expected response $\expresponse$ by processing the previously computed $\challenge$ and storing the output.
The randomness $\randomness$ is input as placeholders which are evaluated when the reply of the provider arrives.
Here, the verifier must ensure that possible influences by the interface channel, such as data encoding, are incorporated.
The verifier also measures the time required to compute $\expresponse$ and adds a certain estimated threshold to respect network latency yielding the maximal allowed time $\timec$ to process the challenge $\challenge$.
Finally, the verifier stores all tuples in the database $\database(\swfamily, \interfaces)$.

\subsubsection{Overcoming Software Version Hierarchy}\label{sec:dbsvhierarchy}
%\CG{HW}
We will now explain how software development and version hierarchies influence the database design.
Recall that two different versions of the same software, i.e., $\version_\sw<\version_{\sw'}$ with $\sw,\sw'\in\swfamily$, might have a very similar latest version history.
That is, they differ from their respective preceding version by nearly the same changes which makes it difficult to distinguish between them a priori.
As explained in Section~\ref{sec:svhierarchies}, this occurs on different software branches that are maintained at the same time.
Howeever, there will be features which have been added to a predecessor of $\sw$, i.e., in the branch of $\sw$, but are not available in $\sw'$, since it exists on a different branch.
Additionally, functions may only exist in the versions between $\sw$ and $\sw'$, but not before or afterwards.
These deprecated functions can not give any information about software that has a lower version number than $\sw$ and a higher version number than $\sw'$.
We will now explain how the verifier copes with these entries in the database and how he finds related branching and deprecated versions.

Coming from the previous step of computing challenges and expected responses, we now have at least one tuple $(\challenge, \expresponse, \timec)_\sw$ for each $\sw\in\swfamily$ in our database $\database$.
To respect and overcome the software properties and version hierarchies mentioned above, we will add additional tests for branch origins and deprecated functions to existing tests if required.
In other words, there can be more than one test for a software version $\sw\in\swfamily$.

Regarding branch detection, this follows from the version history of $\swfamily$ and results in testing for a function of $\fbad$.
That is, if two different software versions $\sw$ and $\sw'$ each have the same change log in respect to their previous version, the verifier determines the next lower branch origin $\widehat{\sw}$ of the higher version, wlog. $\sw'$.
Since we already added a test for this certain software version $\widehat{\sw}$ to $\database$ in the previous step (recall that this was done for all softwares of $\swfamily$), we can simply add this test to the test of $\sw'$.
As an example, let $\dbentry{\sw}$ denote the test for $\sw$ stored in a database $\database$.
Then, if the changes between \emph{(i)} 1.2.2 and $\version_\sw=1.2.3$ and \emph{(ii)} 3.4.4 and $\version_{\sw'}=3.4.5$ are equal with respect to distinguishing sequences for simulation-hard functions, then the test entry in $\database$ is $\dbentry{\sw} = \dbentry{1.2.3}$ for $\sw$, while for $\sw'$ it is $\dbentry{\sw'} = \dbentry{3.4.0} \land \dbentry{1.2.3} =: \dbentry{\widehat{\sw}, \sw}$.
Hence, $\rfps$ learns $\decision$ for either only one or two tests and can act accordingly, e.g., supported by a strategy.
The verifier always wants to detect the branch origin \emph{before} performing other tests, since if this test is not successful, he knows that he can jump immediately to an other branch.
Concatenating tests is not commutative and will be halted as soon as one $\decision=\false$ is output by a single test.
To find $\version'=\version_{\sw'}$ as given in equation~\eqref{eq:bad}, the verifier compares $\sw'$ to each branch that has been active during the release time of $\sw'$. %\sw instead of \sw' (last 2)?
As a final note, a combination of tests as $\dbentry{\sw}=\dbentry{\sw', sw''}$ can also contain further tests, even for $\sw$ itself (which then is \emph{not} a circular reference since only $\sw$-related tests have to be executed).
This is required if multiple consecutive releases on different branches have the same feature difference to its predecessor each.

For deprecated functions, we analogously get a function in $\fugly$ for $\sw$, and add a second test to determine the version range in which the function is available.
That is, we later want to perform a second test \emph{after} testing the specific intrinsic function of $\sw$ to determine its version including the information of deprecated functions.
This way, we know if some $\func\in\fgood(\sw)$ is available via $\dbentry{\sw}$, and can afterwards execute the second check $\dbentry{\sw'}$ for a function $\func'\in\fugly(\version_\sw,\version_{\sw'})$.
This yields $\dbentry{\sw} = \dbentry{\sw, \func} \land \dbentry{\sw', \func'}$.
To find $\version'=\version_{\sw'}$ as given in equation~\eqref{eq:ugly}, the verifier compares the functionality $\func$ of $\sw$ to all adjacent versions to find the version range of $\func$.

Please observe that executing the tests in a given order is possible since the verifier knows during the creation of the database if $\func\in\fbad$ or $\func\in\fugly$ for a functionality $\func$ of $\sw$.
Note that this is not an exclusive or.
If there are two different software versions $\sw$ and $\sw'$ with the same set of simulation-hard intrinsic functions, same branch origin and no deprecated function, i.e. $\fgood$, $\fbad$, and $\fugly$ are the same for both $\sw$ and $\sw'$, we then have $\sw\equiv\sw'$, that is both versions are indistinguishable.

From a practical point of view, the referred tests must not be stored multiple times in the database, for example, a references should point to the entry of that version in the database.
Additionally, the verifier should make sure that each test is only performed once, e.g., by employing $\decisionlog$.
Another important point in practice is that some tests require using features of a software that might not be implemented in the current running version of $\swchallenge$ yet.
That is, if the verifier is testing for some functionality of, e.g., software version 2, but actually the CSP hosts $\swchallenge$ with software version 1, then it might be that $\swchallenge$ is technically not capable of understanding a challenge for software version 2.
Then, the $\swchallenge$ will most likely throw an error (which is not a fingerprinting technique since this holds for \emph{all} versions greater 1).
In this case, software version 1 would be a predecessor that has to be tested before software version 2 is tested.
In other words, software version 1 must be treated as a branch origin of software version 2, i.e., pointing via referer to the according database entry.
As an example, testing for PHP version 7.1.20 only makes technically sense if there was beforehand a test for version 7.0.0.
That is, \RFP tests 7.0.0 and only if this test succeeds, the test for 7.1.20 is performed, i.e., $\dbentry{7.1.20} = \dbentry{7.0.0, 7.1.20}$ (note that this is not a circular reference: the latter test refers to the intrinsic functions of 7.1.20 only).

\subsubsection{Strategy Independence}
Following from the previous steps, we now have a database which consists of entries either with a single test for software versions that can be identified by an intrinsic simulation-hard function only, or with multiple tests if the software version shares common intrinsic functions with a software version from another branch or its intrinsic function is only available for a certain range of software versions.

In this final database design step, the verifier makes sure that the database $\database$ will later on work with any strategy $\strategy\in\strategies$, i.e., $\database$ is independent of any reasonable $\strategy$.
Ultimately, we do not want to restrict the set of possible strategies, i.e., be as flexible as possible.
The reason is that different strategies require a different number of steps to determine the final software version.
This depends on the software that is being audited and the start software version of the strategy.
For the database design it is important that any order of software version tests is applicable.
For example, when a starting with a test for version 5.3.2 it is not sufficient to check for the intrinsic functions of this specific version.
By applying the respective tests referred to in the database, the verifier also needs to check that the software is part of 5.3, and again as part of this he needs to make sure that the major branch 5 is indeed correct.
Otherwise, it is possible that a test erroneously yields a correct output, for example when branching is involved.
Therefore, the verifier has to make sure that any entry point of a strategy leads to the correct result (given the strategy is reasonable).
Furthermore, software may change and branches as well as new versions will be added, requiring to alter existing entries in the database.
Note that if the branch origin tests and deprecated function tests are included correctly in the database for each software version test, no matter where $\rfps$ starts in the database, there will always be a reference to one or more tests, yielding unique results.
Hence, correctness is given for any reasonable strategy.

%The database could be reduced in size if the entry software version is a priori known, since it will result in a static tree with given root and a path depending on the software version tested.
%But software may change, branched get added and grow without rules in general, and the entry point for a strategy may change over time.
%Different software may require different strategies to traverse the software versions in the database, depending also on the a priori version guess of the verifier.
%This could be the highest or lowest possible version for new or old software, respectively, i.e., the verifier could start ''near the final version''.

In conclusion, the main challenges for the verifier regarding creating a database are evaluating the version history, implementing probabilistic properties into intrinsic functions, reference versions in case of branching, deprecated functions, or technical dependencies, and to make the database sound such that any strategy can run on it.
On top, the database should be updated whenever there is a software version $\sw$ added to $\swfamily$, to allow distinguishing $\sw$ from other software versions.
We refer the reader to Section~\ref{sec:expdatabase} for a brief explanation of the database format in practice and the general capabilities of our database implementation with example code included.

\subsection{Strategy Design}\label{sec:strategies}
A strategy $\strategy\in\strategies$, as defined in Definition~\ref{def:strategy}, is constructed by the verifier and consists of two parts:
first, a start entry of a database $\database(\swfamily, \interfaces)$, and second, how to select the next entry in the database based on previous tests and results stored in $\decisionlog$.
On top, a strategy may repeat tests, even the whole strategy itself overall.
This is required, since anomalies in the network between the verifier and tested software can be mitigated by performing time- and function-critical processes multiple times.
For example, a strategy could execute itself after completion of one run in order to test if the CSP behaves differently regarding latency or computation results.

%In order to reduce the influence, e.g., of anomalies in the network or server resulting from a load balancer (or a malicious CSP), tests defined in the database may be performed multiple times instead of only once, which then would be part of the strategy, too.

In order to find a good strategy, we need to recall the three characteristic cases of software version hierarchy from before:
intrinsic simulation-hard functions, version branching, and deprecated functions.
For example, two different versions may share the same intrinsic functions but are part of different branches.
Depending on the outcome of the individual tests, different actions need to follow.
As an example, assume the verifier tests $\swchallenge$ running at the CSP, and he tests against the PHP version $\sw$ with $\version_\sw=7.1.1$.
In the database we have $\dbentry{\sw} = \dbentry{7.1.0}\land\dbentry{7.0.15}$ which gives the following four cases for a strategy:
\emph{(i)}
%if $\sw$ and $\sw'$ share the same intrinsic functions but are on different branches, then it holds
$\decision_{\rfpp(7.1.0)}=\true\land\decision_{\rfpp(7.0.15)}=\true \Rightarrow \version_\swchallenge\geq 7.1.1$;
\emph{(ii)}
%if the intrinsic properties of $\sw'$ are given, but those of $\sw$ are not, then it holds
$\decision_{\rfpp(7.1.0)}=\false\land\decision_{\rfpp(7.0.15)}=\true \Rightarrow \version_\swchallenge\geq 7.0.15$;
\emph{(iii)}
%if the software version branch of $\sw$ is the same as $\swchallenge$, then it holds
$\decision_{\rfpp(7.1.0)}=\true\land\decision_{\rfpp(7.0.15)}=\false \Rightarrow \version_\swchallenge = 7.1.0$; and
\emph{(iv)}
$\decision_{\rfpp(7.1.0)}=\false\land\decision_{\rfpp(7.0.15)}=\false \Rightarrow \version_\swchallenge < \sw_\version \lor \version_\swchallenge < \sw'_\version$.
In each case, the strategy has to decide what to do next, i.e., choose the following software version to be tested.
Hence, the verifier has to keep in mind that combinations of tests will yield certain knowledge about the software version running at the cloud service provider.
This is similar to the linked tests for branching and deprecated functions.
However, in this case the verifier needs to know which version is newer than another one in order to process.
He gains this information from the database, i.e., the version history of $\swfamily$, despite not having a well-defined order supported by the version numbers.

In our reference implementation of \RFP, we implemented the following strategies which work very well for various use cases and at the same time demonstrate the capabilities and possibilities:
\begin{description}
\item[Binary Search (BS)]\ \\
Classical binary search (see e.g. \cite{Cormen:2009:IAT:1614191}) is based on the idea to, first, split a set of ordered items in half and, second, to choose the `correct' half, i.e., the half where $\swsource$ is included.
As already mentioned, there is no trivial way to order the software versions, but due to the structure built beforehand, we know if we are in the correct branch or not after a certain test via $\rfpp$, no matter how we sort all software versions.
Hence, we implemented the ordering over the version numbers, but kept the database structure to possibly jump in a half which was not chosen by the original binary search algorithm (another choice could be to sort by software version release date).
In general, this algorithm is quite efficient in finding the correct software version.
However, if due to software branches an older version number fails a test while a newer one succeeds (e.g., 7.1.0 is fine, while 7.0.15 is not), the binary search algorithm will take a software version that is older than the failed and newrer than the succeeded version -- in this case, this yields a non-existent version and has to be treated accordingly.
\item[Cascading Binary Search (CBS)]\ \\
For this strategy, we perform binary search three times: on the level of \Major, \Minor, and \Patch each in this order.
Depending on the number of different versions and branches, this reveals the target software version quicker than classical binary search.
\item[High To Low (HTL)]\ \\
If the verifier suspects that $\swsource$ at the CSP is running in a relatively recently released version of the highest version branch, then he might employ this strategy.
The first version to test is the highest released version, the next one the second highest released one, and so on.
This is done successively downwards until the correct software version has been found, i.e., $\rfpp$ yielded \true for some $\sw\in\swfamily$.
However, note that a single software version test can reveal more information than just about the currently tested software.
Due to the referred software versions via branching, deprecated functions, and technical dependencies, this strategy can skip a test, since it always selects for the next test the lowest version which has not yet been successfully tested.
This optimistic strategy is less efficient, if the software version tested is one that was not recently published.

Note that there are important details to be aware of:
In the example of PHP, we have $\decision{\rfpp(7.2.9)}=\true \Leftrightarrow \decision{\rfpp(7.1.21, 7.2.0)} = \true$.
Now assume that $\decision{\rfpp(7.2.9)}=\false$.
Then, this strategy \emph{must not} select 7.1.20 as a next version to test if $\decision{\rfpp(7.1.21)}=\false$, but it has to check version 7.2.8.
However, if $\decision{\rfpp(7.1.21)}=\false$ and $\decision{\rfpp(7.2.0)}=\false$, then in fact $\rfpp(7.1.20)$ must be the next test.
\item[Low To High (LTH)]\ \\
This strategy works analogously to High To Low, but starts from the oldest (e.g., lowest when ordered by version number) software version of $\swfamily$ ever released.
\item[Highest Major Step Up (HMSU)]\ \\
The entry point in the database for this strategy is the software version with the highest \Major, but lowest \Minor and \Patch values each.
The strategy is optimistic in the sense that it jumps directly to the next higher \Minor branch, if a test was successful.
However, if a test was not successful, it will test the next higher \Patch value.
If both the test for \Minor and \Patch fail, a lower \Major will be tested.
This, of course, also holds, if the very first test fails.
HMSU is well suited for software families which have a low amount of different \Major values and when the CSP is not running the most recent version at the same time.
\end{description}
In conclusion, strategies define how the software versions in a database are traversed.
The implemented strategies demonstrate just a small part of the bandwidth possible for the verifier to create a strategy.
We will compare the listed strategies to each other in Section~\ref{sec:experiments}, when testing a real-world scenario as part of the experiments, and give a thorough description of a single strategy (CBS).
However, analyzing the runtime of the strategies, e.g., average and worst case, is out of scope of this paper.

\subsection{Implementation Modularity of the \RFP Framework}
The structure and design of the \RFP scheme $\rfps$ allows to have a modular implementation framework.
That is, $\rfps$ consists of various components, such as databases and strategies, which encapsulate different tasks.
In fact, the different building blocks of $\rfps$ interact with each other, but are independent developed components.
This yields the following modules:
\begin{enumerate}
\item Databases $\databases$, where a single database includes the pairs of challenges $\challenge$ and expected responses $\expresponse$ together with a time limit $\timec$ for each software $\sw\in\swfamily$ (and possibly referrers to other entries, see Section~\ref{sec:svhierarchies}).
Fingerprinting techniques, which describe how to evaluate a certain functionality of $\sw$, can be developed independently of all other components, however, they are implemented as part of a database entry.
\item Interfaces $\interfaces$, which represent communication channels and can be freely chosen for each challenge and each (expected) response.
An interface may take as input authentication data negotiated with the software provider $\prover$.
\item Strategies $\strategies$, dictating how to choose the next software version $\sw\in\swfamily$ to be audited as part of $\rfpp$, depending on earlier test results $\decisionlog$.
\item A randomness source $\randsource$, which is applied during runtime yielding an auxiliary input to each challenge.
\end{enumerate}
This flexibility results in a very adaptive nature of \RFP.
The verifier is independent of the software family and software complexity (e.g., a forum runs on PHP, which runs on Apache, which runs on Ubuntu, etc.).
Furthermore, due to the modularization, the verifier can construct databases independently of the software provider, as authentication data can be input as a parameter.
Strategies are not only independent of the software family, but also of the database; latter may be an input parameter.
Consequently, a \RFP scheme is quickly extensible and a verifier can outsource the design of databases and strategies, exchange them efficiently for each software family if required and may even share them with further parties.
Finally, note that \RFP is not bound to any infrastructure or programming language, as long as it can be executed.

\subsection{Experiments}\label{sec:experiments}
After having defined \RFP and its scheme $\rfps$ we want to test it in a real-world setting.
Hence, we performed experiments with actual service providers and will also come back to the attack example of Section~\ref{sec:practicalattack}.
According to the database and strategy design explained above, we created multiple databases and strategies.
The databases include tests for each software family PHP, MySQL, and WordPress, while the strategies are the ones mentioned in Section~\ref{sec:strategies} and can be applied to any of the databases.

\subsubsection{Experiment Database}\label{sec:expdatabase}
For storing the databases, we used the lightweight data-interchange format JavaScript Object Notation (JSON, \cite{rfc8259}), consisting in general of key-value pairs and ordered lists.
A database is created by first fetching all available software version names from the developer of the software family.
Next, some metadata is set, such as the timestamp for creation and last change (this info makes updating the database easier later on), the name of the software family, e.g., PHP, default values for software specific values, e.g., challenge and expected response appending start- and endstrings or default latency, default randomized variable type and format, and which interfaces should be applied by default, i.e., if none are specified for a specific test.
See Appendix~\ref{sec:exampledbmeta} for an example configuration.
Next, the tests for each version of the software family is added according to the database design presented in Section~\ref{sec:dbdesign}.
Each software version may consist of a set of intrinsic function tests, a set of branching version references (including technical dependencies as explained in Section~\ref{sec:dbsvhierarchy}), and a test for deprecated functions.

As an example, the test for PHP 7.2.12 can be given as seen in Listing~\ref{lst:php7.2.12}.
Hence, to check if at least PHP 7.2.12 is running as the provided software, versions 7.2.0 and 7.1.24 are tested, since 7.2.12 and 7.1.24 share the same intrinsic functions compared to their previous versions each, and 7.2.12 lives on branch 7.2.
\begin{lstlisting}[basicstyle=\footnotesize, numbers=left, language=bash, frame=lines, framesep=10pt, numbersep=-8pt, caption={\RFP example database test entry for PHP 7.2.12.}, label=lst:php7.2.12]
    "7.2.12": {
      "test": {
        "branching": {
          "7.2.0": "1",
          "7.1.24": "1"
        }
      }
    }
\end{lstlisting}

We will now give an example for a test based on an intrinsic function.
Assume that we send challenges via FTP (i.e., storing a file) and receive responses over HTTP (access an URL whose content is provided by PHP and a webserver).
Then, the test for PHP 7.2.0 may be given as in Listing~\ref{lst:php7.2.0}.
First, a variable \texttt{ax} is defined as a random integer\footnote{We implemented also randomized formats such as strings, binary values, version information, or directories and files.} between $1$ and $999999999$.
Next, the challenge payload calls the function \texttt{unserialize}, which unserializes the randomized string \texttt{d:\#ax\#e++2}, which has, by intention, not the correct format for a serialized float number.
For example, \texttt{d:3e+2} represents $3\cdot 10^2 = 300$.
Before PHP 7.2.0, the software replies with \texttt{float(3)} in the given example, which is not compliant to the definition and hence wrong.
With the introduction of PHP 7.2.0, the response for such a string consists correctly solely of a (variable) error message.
With this knowledge, we can test if the payload returns \false, denoting that at least PHP 7.2.0 is running.\footnote{For this example we disabled error reporting, hence the output consist only of the boolean value \false.}
We use the function \texttt{var\_dump} to transform the output of any argument into a string.
The expected payload contains the value that we are expected to get as a response from the CSP.
Note that we are using the fingerprinting technique \fterrorhandling with the intrinsic simulation-hard function \texttt{unserialize} to determine if at least version 7.2.0 is running here.
Also, recall that due to the randomization and time-constrain, the software provider can not pre-compute, cache, or forward the challenge and has, in particular, to execute the challenge in the actual target environment.
\begin{lstlisting}[basicstyle=\footnotesize, numbers=left, language=bash, frame=lines, framesep=10pt, numbersep=-8pt, caption={\RFP example database test entry for PHP 7.2.0.}, label=lst:php7.2.0]
    "7.2.0": {
      "test": {
        "variables": {
          "ax": {
            "format": "integer",
            "min": 1,
            "max": 999999999
          }
        },
        "challenge": {
          "payload": 
            "var_dump(@unserialize('d:#ax#e++2;'));"
        },
        "expect": {
          "payload": "bool(false)\n"
        }
      }
    }
\end{lstlisting}
Note that, in comparison to \RFP, the often used fingerprinting techniques of Class~1 merely query a version API, e.g., \texttt{phpversion()} in the case of PHP, \texttt{SELECT VERSION()} for most SQL databases, or a banner for command-line software.
Recall that the response can then be freely chosen by the (malicious) service provider.

\subsubsection{Defying the Practical Attack}
We implemented the \RFP scheme in PHP, allowing us to port it quickly to any other system.
Of course, any high-level programming language will also work perfectly fine.
Recall the practical attack on fingerprinting techniques from Section~\ref{sec:practicalattack}:
we set up a malicious server that is running PHP 7.1.1, but on challenging any version function, the server replies with \texttt{20.0.85-car}.
The tested version scanners were convinced that this non-existent software version is indeed running.

We will now leverage \RFP to determine the software version of this malicious server and compare the outcome to other tools.
For easier use, we implemented an \RFP-Wizard, which is part of \RFP and can cover technical details such as the authentication and network setup as well as location of the database and service provider.
The wizard is also available as API, e.g., via REST.
To begin, we first specify the database $\database$ with software family $\swfamily=\text{PHP}$ (378 versions, lowest is 4.0b1, and highest is 7.3.0rc5) and interfaces $\interfaces=\{\text{FTP},\text{HTTP}\}$, and next the credentials for the challenges sent via FTP and responses received over HTTP.
Finally, we provide \RFP with an a priori customer target version claim, for example version 7.3.0.
In fact, this is the information the customer of the service provider usually gets from their customer web interface.
In other words, this is the advertised or in an SLA agreed on software version the provider has promised to run.
As mentioned in Section~\ref{sec:basicidea}, we now have a \RSI protocol $\rsip$ with \RFP as a building block.
We perform \RFP with the strategies Cascading Binary Search and Highest Major Step Up to compare both.
The test procedure results are given in Table~\ref{tbl:malicioustest}, where ``Version'' denotes the PHP version tested, ``Result'' is the outcome of the test -- either \true (\cmark) or \false (\xmark) --, and ``Testorder'' is the sequence in which the tests have been performed, i.e., the distinguishing sequence.
While the strategies traverse different paths through the possible software versions, both determine correctly 7.1.1 as the software version running at the service provider.
Hence, \RFP overcomes the problems of existing tools and can detect software versions even in the presence of a malicious service provider.
More formally, \RFP yields $\version_\swsource = \version_{\fsmsource} = \text{7.1.1}$, $\version_\swchallenge = \version_{\fsmchallenge}=\text{20.9.85-car}$, and $\version_\swtarget = \version_{\fsmtarget}=\text{7.3.0}$, while other tools yield erroneously $\version_{\fsmsource}=\version_{\fsmchallenge}$.
To complete an \RSI, the verifier has to check if the CSP is compliant to the SLA, i.e., if it holds that $\swsource=\swtarget$ (which is not the case here).
\begin{table}
\centering
\begin{tabular}{ccc|ccc}
\toprule
\multicolumn{3}{c|}{\textbf{Cascading Binary Search}} & \multicolumn{3}{c}{\textbf{Highest Major Step Up}}\\
\textbf{Version} & \textbf{Result} & \textbf{Testorder} & \textbf{Version} & \textbf{Result} & \textbf{Testorder} \\
\midrule
5.0.0b1 & \cmark & 1 & 7.0.0  & \xmark & 1 \\
7.0.0   & \cmark & 2 & 7.1.0  & \cmark & 2 \\
7.2.0   & \xmark & 3 & 7.2.0  & \xmark & 3 \\
7.1.0   & \cmark & 4 & 7.0.15 & \cmark & 4 \\
7.0.26  & \xmark & 5 & \textbf{7.1.1}  & \cmark & 5 \\
7.0.22  & \xmark & 6 & 7.1.2  & \xmark & 6 \\
7.1.2   & \xmark & 7 & & &\\
7.0.15  & \cmark & 8 & & &\\
\textbf{7.1.1}   & \cmark & 9 & & &\\
\bottomrule
\end{tabular}
\caption{\RFP test procedure results of a malicious PHP service provider with the faked software version number \texttt{20.9.85-car}.
\RFP determines the correct version \texttt{7.1.1}, employing the strategies CBS and HMSU each for comparison.}
\label{tbl:malicioustest}
\end{table}

\subsubsection{Evaluating Real-World Software Providers}
We tested various software providers regarding their promises of the running software version using \RFP, e.g., Hetzner \cite{Hetzner}, Serverprofis \cite{Serverprofis}, or Strato \cite{Strato}.
As a result, while no provider was really wrong, some of them promised to have ``the newest version'' installed, but did not provide a \Patch value of the version number running in the customer web interface.
This may be due to save content update time, since this reduces the text changes the providers have to make each time a new version is being released.
However, at each point in time, the most recent version for a given \Major and \Minor branch is always clearly defined at the tested providers.
We found that all but one of the providers indeed hat the most recent version installed.
The other was between a single up to five \Patch versions behind.
In contrast, for MySQL, no version number was provided in the customer web interface, but it got advertised by the \Major version, leaving open both \Minor and \Patch versions.
Not updating to the newest version will result in less secure software (except some edge cases).
Hence, it is crucial to keep software up to date.

We will now use \RFP to determine the most recent version of PHP 7.2 running at the platform as a service provider `Serverprofis'.
In order to demonstrate how different strategies analyze the same software, we are performing this procedure for each strategy mentioned in Section~\ref{sec:strategies}, except Low To High.
The result is presented in Table~\ref{tbl:strattest}, where ``R.'' is short for ``Result'', and the tests are listed in the order they have been performed.
At the end, \RFP proves that indeed the most current version of PHP 7.2 is running at the moment of testing, i.e., 7.2.14.
\begin{table}
\centering
\begin{tabular}{cc|cc|cc|cc}
\toprule
\multicolumn{2}{c|}{\textbf{BS}} & \multicolumn{2}{c|}{\textbf{CBS}} & \multicolumn{2}{c|}{\textbf{HTL}} & \multicolumn{2}{c}{\textbf{HMSU}}\\
\textbf{Version} & \textbf{R.} & \textbf{Version} & \textbf{R.} & \textbf{Version} & \textbf{R.} & \textbf{Version} & \textbf{R.} \\
\midrule
5.2.0    & \cmark & 
5.0.0b1  & \cmark & 
7.3.0rc4 & \xmark &
7.0.0    & \cmark \\
7.1.0    & \cmark & 
7.0.0    & \cmark & 
\textbf{7.2.14} & \cmark & 
7.1.0    & \cmark \\
7.0.15   & \cmark & 
7.2.0    & \cmark &
         &        &
7.2.0    & \cmark \\
7.1.1    & \cmark & 
7.3.0rc4 & \xmark &
         &        &
7.3.0rc4 & \xmark \\
7.1.21   & \cmark & 
7.1.21   & \cmark &
         &        & 
7.1.13   & \cmark \\
7.2.0    & \cmark & 
7.2.9    & \cmark &
         &        &
7.2.1    & \cmark \\
7.0.0    & \cmark & 
\textbf{7.2.14} & \cmark & 
         &        & 
7.1.14   & \cmark \\
7.1.20   & \cmark & 
         &        & 
         &        & 
7.2.2    & \cmark \\
7.2.8    & \cmark & 
         &        & 
         &        &
7.1.20   & \cmark \\
\textbf{7.2.14} & \cmark & 
         &        & 
         &        &
7.2.8    & \cmark \\
7.3.0rc4 & \xmark & 
         &        & 
         &        &
7.1.21   & \cmark \\
         &        & 
         &        & 
         &        & 
7.2.9    & \cmark \\
         &        & 
         &        & 
         &        & 
7.2.11   & \cmark \\
         &        & 
         &        & 
         &        & 
\textbf{7.2.14}   & \cmark \\
\bottomrule
\end{tabular}
\caption{\RFP proves that the software provider `Serverprofis' has indeed running the latest version of PHP 7.2 at a certain point in time.
The table shows a comparison of four different strategies employed in this \RFP test: BS, CBS, HTL, and HMSU.}
\label{tbl:strattest}
\end{table}

Finally, we want to explain in detail one strategy, such that the reader knows why the versions have been tested in the order presented in Table~\ref{tbl:strattest}.
Hence, we will briefly describe Cascading Binary Search (CBS).
The first goal of this strategy is to determine the correct \Major version.
The database contains 4, 5, and 7 as candidates, and CBS starts in the middle with \Major version 5 and the lowest available version thereof, i.e., 5.0.0b1.
In this case, we get \true as a result, hence, the tested software might have with a higher \Major version and we test for version 7, i.e., 7.0.0, which results in \true, too.
Next, we want to find the correct \Minor version.
For PHP 7, we have the following \Minor versions in the database: 0, 1, 2, and 3.
Choosing again the (optimistic, i.e., rounding up) middle, the test $\rfpp(7.2.0)$ results in \true.
Note that we do not have to check 7.0.0 again.
Next, however, testing for \Minor version 3 fails.
We now are looking for the final part, determining \Patch out of the possible values in our database, i.e., 1, 2, 8, 9, 11, and 14.
We want to show here that without a perfectly filled database, \RFP still manages to find the correct version, as long as it is included.
Observe that \Patch value 0 was already tested.
The last two steps are successfully testing for \Patch version 9 and 14.
Since there is no version of PHP between 7.2.14 and 7.3.x, we do not need to conduct further tests and CBS can output the final, correct software version as result: 7.2.14.

\section{Extensions of \RFP}
In this section we describe two extensions of \RFP: fingerprinting hardware and outsourcing the burden of creating databases and strategies of the verifier to an external auditor.

While a software provider simply might act as a verifier in the \RFP scheme $\rfps$ in order to audit himself, this is not an extension of \RFP.

\subsection{Fingerprinting SD Cards}
Until now, we presented \RFP as a method to determine the identity of software, i.e., the software's correct version.
However, the same concept behind the \RFP scheme can also be applied to verify the correctness of hardware features.

One case is the overall available storage on an SD card (Secure Digital Memory Card \cite{sdcard}).
The maximum storage available is effectively just represented by a string of characters which is then consumed by other applications.
Usually, this string is not checked by any operating systems and is ``believed'' to be correct as it is stored on a certain, in general non-accessible part of the card.
This scenario is very similar to software, whose version number string is also being ``believed'' to be correct without any checks.

The idea is now to extend \RFP by testing for available space on the SD card until we find the maximum amount possible (or at least a close range thereof).
To do so, we set the \RFP database to not contain version numbers, but various magnitudes of storage sizes, e.g., 1 kilo byte (kB), 10kB, 100kB, and so on.
For each challenge, the precise value of the storage size is chosen randomly (in the range of the chosen magnitude), as well as the content.
Then, the data is transmitted to the previously erased SD card over an interface provided by the operating system or network.
As in original \RFP, various strategies may be applied, e.g., binary search (which is easy here, since a clear order on the target space exists -- storage space).
At some point in time, the SD card will stop taking data because it is full.
Then, the verifier has learned the maximum capacity of the SD card.

This extension of \RFP is especially of interest, since there are many SD cards which claim to have a higher storage capacity than they really possess.
The motivation behind such products are manufacturers who buy cheap low-storage SD cards and replace only the string with the available space value in it.
This, is quite easy given a certain interface, and allows selling cards of a Terabyte which identify as such, but can only store a few Gigabyte.
The customer most often will only notice this fact when it's too late.
We want to point out that there exists software, that determines the size of a storage object reliably in different ways, if used correctly (e.g., formatting software), \RFP complements these tools.

\subsection{Outsourced \RFP}
We have presented \RFP as a two-party protocol between a verifier $\verifier$ and a (cloud) software provider $\provider$.
As we have seen in Sections~\ref{sec:dbdesign} and \ref{sec:strategies}, the verifier has to construct a database for each software family $\swfamily$, employ randomness, and implement strategies to perform \RFP.
However, a usual customer of the provider may not have the required resources, i.e., time, money, computing, or storage power.
Hence, we introduce a third party to take away the burden of the customer: the auditor $\auditor$.
The goal of the auditor is to check on behalf of the customer if the provider behaves correctly, and in the case that something is not working as intended, i.e., as defined in according SLA, he informs the customer.
In practice, however, the provider may detect based on the interface if a \RFP challenge comes from a customer or from $\auditor$, and act differently.
That is, $\provider$ will use a certain server instance to reply to challenges of $\auditor$, where this server has installed the most recent software version of $\swfamily$.
Note that this is not proxy forwarding since this server can co-exist directly beside (virtual or non-virtual) the server the customer is using.
In other words, $\auditor$ can not check $\provider$ directly or independently of the customer.
Observe that this is different for functionality than, e.g., storage.
However, the auditor can nonetheless take care of the expensive parts of \RFP, i.e., building database and strategy.

We are now going to extend \RFP by adding the auditor party $\auditor$ and new security definitions and measures to \RFP.
We dub this scheme \emph{Outsourced \RFP (\ORFP)}, since the customer outsources above mentioned tasks to the auditor.
This is similar to the ideas of Outsourced Proofs of Retrievability \cite{DBLP:conf/ccs/ArmknechtBKLR14}, where proofs of retrievability are outsourced to an auditor to verify outsourced data availability.
Since three parties participate in an \ORFP, more than one party may be malicious, and, additionally, two parties may collude to cheat the third one.
Hence, each party must be able to prove that it did its job correctly at any given moment in time.
%For example, a malicious auditor may share his secret keys and the database with the service provider, allowing them to answer any challenge, since the expected response is always known.
That is, a customer must be able to audit the auditor to check if he did stick to the protocol while at the same time the auditor must be able to prove that he did everything correctly, i.e., prove his liability.

\begin{figure}
\centering
\includegraphics[width=\columnwidth]{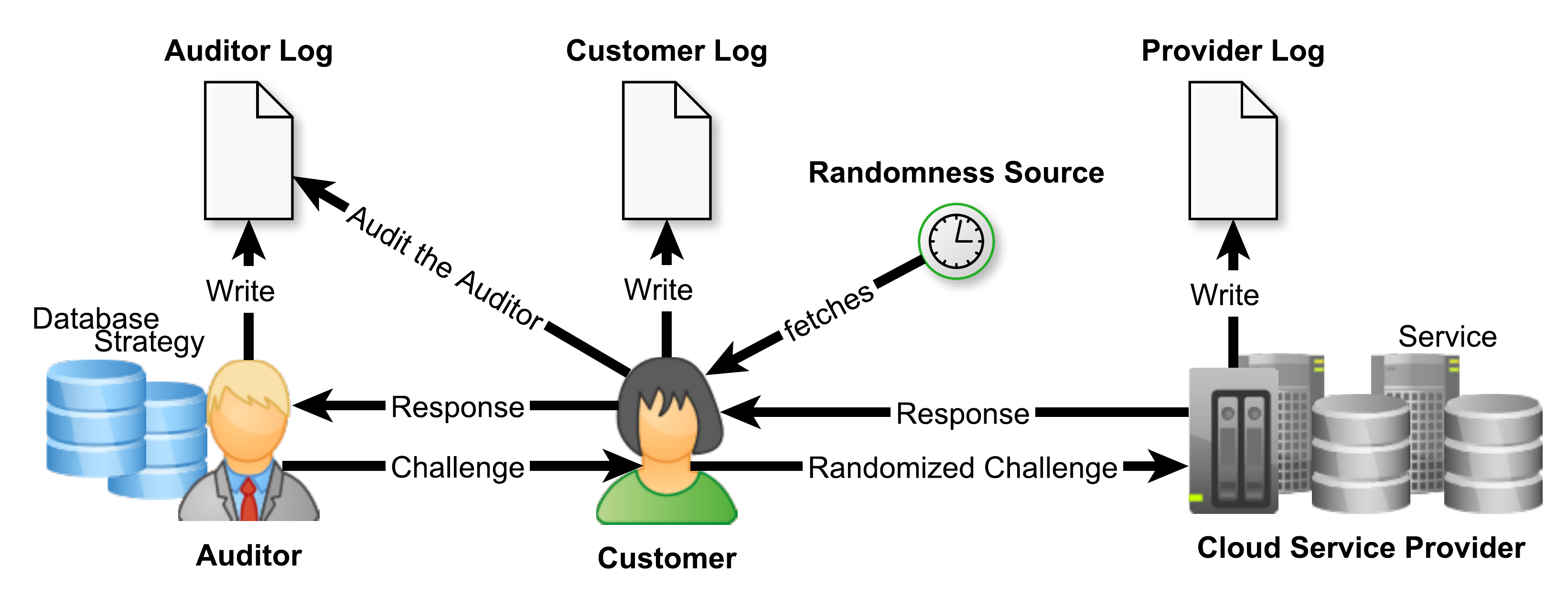}
\caption{Sketch of the \ORFP scheme consisting of the customer, auditor, and (cloud) service provider.
Each party generates values and stores signed values of all signed in- and outputs in a respective log file.}
\label{fig:orfp}
\end{figure}

The core idea of \ORFP is twofold: (i) the auditor generates the database and strategies for a software family (or multiple thereof) while the customer inputs the randomness into the challenge and sends it to the service provider; and (ii) each party keeps a log of all input and output values together with their respective timestamps signed by the generating party.
The first point shifts the main burden of \RFP from the customer to the auditor, while the second point allows to protect against malicious parties and to prove liability.
The \ORFP scheme is depicted in Figure~\ref{fig:orfp}.

%\CG{why is the user still involved?
%a) randomnenss
%b) technical reason: IF the auditor connects directly to the CSP, the CSP might answer differently Hencce, the user must answer. Compared to storage, the functionality can be achieved by having installed a certain software version at a certain time, which could for example be solved by software containers. However, in storage, it's different, since either the full outsourced data is stored or it is not, the CSP especially cannot delete information and later on recreate it. Note that this is possible for functionality.}

\subsubsection{\ORFP Scheme Description}
The \ORFP scheme begins with a \emph{setup phase} which corresponds to the procedure $\procsetup$ of \RFP, where private and public keys are generated and distributed for each party.
Then, the customer/user $\user$ sends a interface set $\interfaces$ to the auditor $\auditor$ together with the software family $\swfamily$.
Here, $\interfaces$ is chosen such that the auditor does not learn the credentials of the user but latter will use it to send and receive data from the service provider $\prover$, e.g., FTP and HTTP.
The user may also send a software version $\version_\swtarget$ representing the software version the provider should have running, otherwise the auditor can set $\version_\swtarget$ to the latest version of $\swfamily$.
Furthermore, $\user$ initializes a randomness source $\randsource$.
In the final part of the setup phase the auditor computes the database $\database(\swfamily, \interfaces)$, strategies $\strategies$, and each party creates a respective log file $\auditlog_\user$, $\auditlog_\auditor$, or $\auditlog_\prover$.

Next, the \emph{audit phase} extends the \RFP procedure $\proctest$, where $\auditor$ begins by choosing a strategy $\strategy\in\strategies$ which is used for the whole phase.
Next, $\auditor$ selects a challenge $\challenge$ from the database $\database$ according to $\strategy$ and sends it to $\user$ together with a signature $S_1 :=\signature{\auditor}{\challenge, \timenow_1}$ where $\timenow_1$ is the current time.
Please observe that we omit the private key inputs of the signature here.
Then, $\user$ induces randomness $\randomness\in\randsource$ into the challenge producing an altered challenge $\challenge'$.
He also computes the signature $S_2 := \signature{\user}{\randomness, \timenow_2, S_1}$ and sends it together with $\challenge'$ to $\prover$.
The $\prover$ then performs the computation of $\challenge'$ resulting in the response $\response$, determines $\timenow_3$, and computes the signature $S_3 := signature{\prover}{\response,\timenow_3, S_2}$.
He adds to his log $\auditlog_\prover$ the values
$(\challenge', \response, \timenow_3, S_2, S_3)$.
Next, the client fetches $\response$, $\timenow_3$, and $S_3$, computes $S_4 := \signature{\user}{\timenow_4}$ based on the current time $\timenow_4$, and forwards these values together with $\randomness$, $\timenow_2$, and $S_2$ to $\auditor$.
He furthermore adds to his log $\auditlog_\user$ the values
$(\challenge, \randomness, \response, \timenow_2, \timenow_3, \timenow_4, S_2, S_3, S_4).$
In the last step, $\auditor$ validates $\response$ dependent on both $\challenge$ and $\randomness$, and then decides to either start a next test (according to the protocol $\rfpp$ and strategy $\strategy$) or, when no further tests shall be done, he outputs a candidate set $\candidateset$ via the procedure $\procoutput$.
If $\version_\sw\not\in\candidateset$, the auditor informs $\user$.
The auditor also updates his log $\auditlog_\auditor$ by adding the entry
\[ (\challenge, \randomness, \response, \timenow_1, \timenow_2, \timenow_3, \timenow_4, S_1, S_2, S_3, S_4). \]

\subsubsection{\ORFP Security Outline}
The properties correctness and soundness of \ORFP follow from $\rfpp$-correctness and $\rfpp$-soundness, since \ORFP basically incorporates the \RFP scheme $\rfps$.
For example, this can be seen by grouping $\auditor$ and $\user$ as the party verifier.

We begin by showing that $\user$ must draw $\randomness\in_R\randsource$.
An \ORFP scheme employs randomness, as \RFP, to prevent pre-computation or caching of the correct response by $\prover$.
Hence, the value $\randomness$ must be drawn by $\user$ or $\auditor$.
First, let's suppose $\auditor$ chooses the randomness value.
He might collude with $\prover$ and communicate with him the expected response such that $\prover$ does not need to have any software installed at all.
In comparison to OPOR, once deleted data can not be un-deleted and the answer depends on user chosen values, however, certain functionality can be cheated any time here and $\prover$ has no data depending on $\user$ in \RFP.
If $\auditor$ colludes with $\user$ to fool $\prover$, they can challenge for some randomness value and afterwards pretend that they have asked for a different randomness value.
However, since the CSP stores the challenge in his audit log $\auditlog_\prover$, he can provably show which challenge was actually sent to him.
Overall, neither $\prover$ nor $\auditor$ can securely input the randomness value into \ORFP.
Next, we investigate the user $\user$ selecting the randomness value.
If he colludes with $\prover$, they do not have the correct response to send to $\auditor$ since $\user$ may choose the same randomness each time such that pre-computation or caching might be possible.
However, they need to send this value to $\auditor$ such that he can compute the correct expected response.
However, if $\user$ sends each time the same value to $\auditor$, he will be alarmed.
At the end, they are not able to fool $\auditor$, since he can compute the expected response correctly based on his log $\auditlog_\auditor$ and hence show that he did his job correctly.
On the other hand, if $\user$ colludes with $\auditor$, again the audit log $\auditlog_\prover$ allows $\prover$ to show that he acted according to the protocol.
In conclusion, the user $\user$ draws the value $\randomness$ from $\randsource$.

The randomness source $\randsource$ itself must not be a public external randomness source (as in OPOR), since $\auditor$ has also access to this source and could pre-compute the randomness values to share the expected responses with $\prover$, fooling $\user$.
However, $\user$ might employ any (pseudo-)randomness only known to him or he involves a private key in determining the randomness value.

We will now outline the third component of \ORFP security: liability of each party.
That is, a party can prove that it behaved correctly according to the \ORFP protocol.
Furthermore, other parties need to be able to verify this, e.g., the user must be able to audit the auditor.
The reader is referred to OPOR, where thorough definition and security analysis of liability is given.
However, the scheme implementation is different to \ORFP.

First, assume the auditor $\auditor$ wants to prove his liability.
He can do so due to the audit log $\auditlog_\auditor$, which consists of all values of previous \ORFP runs that are required for the auditor to compute his output.
Since each value is signed by the generating party, the auditor can prove that he did his job correctly.
Given the case that the auditor has not correctly generated the challenges and expected responses as part of the database, the user $\user$ can verify this by comparing the inputs and outputs of $\auditlog_\auditor$ and $\auditlog_\user$.
By performing \RFP onto the data of the logs, $\user$ must come to the same output as $\auditor$, otherwise liability is not given.

The same procedure can be applied for any liability proof of a party, since each party holds an audit file with values signed by the other parties.
Note that this is applicable in both ways: a party can show that it has done everything correctly and that another party can verify this.
In other words, if a version claim does not succeed, the user can check if the mistake was by the service provider (wrong response) or the auditor (wrong challenge/expected response).
Also, each party signs the values sent to the succeeding party in order to manifest that these values were generated correctly, i.e., they may be tested against a neutral server.

For future research the auditor may fetch the response of the service provider directly.
Here, the main challenge is that the customer and auditor must have synchronized clocks for precise time measurement.

\section{Related Work}\label{sec:relatedwork}
In the following, we name for different scenarios respective approaches the most prominent related work and explain the relation to \RFPlong.
We do not include discovery scanner who try to determine all running services on a system.
In Table~\ref{tbl:relwork}, Appendix~\ref{sec:relworktable}, we give an overview at a glance of different tools and their employed fingerprinting techniques.
Many of the tools presented advertise two different modes of fingerprinting techniques: normal and aggressive.
Usually, normal translates to \ftversionclaim or \ftversionapi and aggressive to a second guess by using another technique, often \ftfilestructure.
None of the analyzed tools is using a strategy to detect a version number as RFP does.
Except of one, all tools rely on trusting fixed strings received from the tested software.
Depending on the task of the tool, fingerprinting is only one step in a chain of actions, e.g., determining a version number in order to select certain attack vectors for a given service like a network vulnerability test that checks a service for the presence of specific security problems.

%\subsection{Penetration Testing} 
\emph{Penetration testing} aims to check if a system can be effectively attacked.
Existing penetration test software relies on knowing the software and version number of the victim in order to perform the tests efficiently.
Hence it may follow up after employing \RFP.
In fact, implementing \RFP as a module is an interesting task for the future.
%%%%%%%%%%%%%%%%%%%
A well known framework in this context is Metasploit\cite{metasploit}.
However, by default it is relying on software version APIs to detect software version in the first place.
For example, this holds for testing VMware Horizon\cite{horizon} or PostgreSQL\cite{postgresql} via Metasploit.
%%%%%%%%%%%%%%%%%%%
Armitage\cite{armitage} is a graphical cyber attack management tool for Metasploit which visualizes targets, exploits, and provides advanced capabilities of the framework.
It also uses the target software's API to determine the software version.
%%%%%%%%%%%%%%%%%%%
The PenTesters Framework\cite{pentestersframework} is a tool to manage penetration testing tools, but does not employ own fingerprinting techniques.

%\subsection{Vulnerability Scanners} 
\emph{Vulnerability scanners} search for software and network vulnerabilities, such as known CVEs\cite{cve}, unsecure configurations, or missing updates.
Usually, they produce a list of technical security flaws and can be integrated into SIEM (security information and event management) software.
%Given a software and its version, or sometimes also determined by a scanner, they then also cope with the attack vectors.
%%%%%%%%%%%%%%%%%%%
The Burp suite\cite{burp} is a prominent tool for security testing of web applications.
While it has the capabilities of testing public functions in general, it natively relies on the target software's version API when determining its version.
Again, \RFP could be implemented as an extension in the future to solve this problem.
%%%%%%%%%%%%%%%%%%%
The Burp extension Software Version Reporter\cite{softwareversionreporter} scans passively for applications revealing software version numbers.
The software version is determined by the version claim made by the target's service provider.
%%%%%%%%%%%%%%%%%%%
Nessus\cite{nessus} is a network and vulnerability scanner for operating systems that acts mainly as a port scanner, but also scans default passwords or configurations.
That is, specific server software must run on the target server which clients can connect to in order to perform scans.
However, \ftversionclaim and \ftversionapi of the target are used to determine software version numbers.% and it fulfills a different system model than RFP does, i.e., focusing on other software stack layers.
%%%%%%%%%%%%%%%%%%%
Greenbone/OpenVAS\cite{openvas} is a fork of Nessus and is developed on its own over the last decade.
It consists of vulnerability management and vulnerability scanners, focusing on network scanner, operating system, and configurations.
Software versions are determined by employing \ftversionapi.
%Additionally, it is not suited for web- or cloud-services.
%%%%%%%%%%%%%%%%%%%
Another tool in this category is nmap\cite{nmap}.
It scans ports, detects active firewalls, consists of a Remote-Procedure-Call scanner, but does not scan web- or cloud-services, and relies on \ftversionclaim or \ftversionapi provided by the target entity.
However, nmap sends specific crafted TCP packets to provoke error messages which in turn allows for OS and network version detection, if not handled correctly by the target server and software.
%%%%%%%%%%%%%%%%%%%
w3af\cite{w3af} is a scanner for vulnerabilities using a modular structure with the goal to identify, audit, and exploit vulnerabilities.
While it performs attacks like cross-site scripting and SQL injection, the software version is not determined but given by the user.
%Hence, RFP could be a tool you want to use before executing w3af.
%%%%%%%%%%%%%%%%%%%
Both tools Web Application Scanning\cite{webapplicationscanning} and FreeScan\cite{freescan} by Qualys scan computers, websites, and apps for vulnerabilities.
To determine a version number of a software, they rely on \ftversionclaim and \ftversionapi.
%%%%%%%%%%%%%%%%%%%
InsightVM\cite{insightvm} is a platform-based vulnerability scanner and management tool by Rapid7 and employs \ftversionapi to determine version numbers.
%%%%%%%%%%%%%%%%%%%
Nexpose\cite{nexpose} is a similar tool but for on-premise use.
%%%%%%%%%%%%%%%%%%%
OWASP Nettacker\cite{nettacker} is a network information gathering vulnerability scanner and relies on \ftversionapi for fingerprinting.

%\subsection{Version Scanners}
The goal of \emph{version scanners} is to determine the version of multiple softwares employed at once in a web- or cloud-service, for example webserver and CMS.
However, to the best of our knowledge, all of them rely on \ftversionclaim and \ftfilestructure.
%%%%%%%%%%%%%%%%%%%
Popular version scanners are wappalyzer\cite{wappalyzer} (identifies technologies on websites including their version numbers), WhatRuns \cite{whatruns} (also identifies technologies and their version numbers on websites), WhatWeb\cite{whatweb} (a ``next generation web scanner'' to determine employed libraries, software, and versions), and  Guess\cite{guess} (detects CMS, frameworks, webserver, libraries, and versions).
%%%%%%%%%%%%%%%%%%%
Blind Elephant\cite{blindelephant} by Qualys uses \fthashdigest to fingerprint web applications.
%%%%%%%%%%%%%%%%%%%
Similar tools are WAFP\cite{wafp} and Static File Fingerprinting tool\cite{staticfilefingerprinting} by Sucuri, both also comparing checksums of static files.

%\subsection{Content Management System Scanners}
\emph{CMS scanners} are specialized on detecting versions, vulnerabilities, and plugins of content management systems (CMS).
%Remember that the (malicious) service provider is allowed compute any version claim or file as long as it costs less effort than updating the software.
%%%%%%%%%%%%%%%%%%%
The tool What CMS?\cite{whatcms} detects about 300 different content management systems, however, the CMS version relies on \ftversionclaim and file structure (\ftfilestructure).
%%%%%%%%%%%%%%%%%%%
Similarly, wpscan\cite{wpscan} is a security scanner for WordPress, but identifies the version number based on \ftversionclaim and \ftfilestructure.
%%%%%%%%%%%%%%%%%%%
The following four tools rely only on static file content (\ftfilestructure) on the service host:
Plecost\cite{plecost} (WordPress fingerprinting tool),
WPSeku\cite{wpseku} (WordPress security scanner),
JoomScan\cite{joomscan} (OWASP Joomla vulnerability scanner project), and
CMSScanner\cite{cmsscanner} (general CMS scanner framework).
%%%%%%%%%%%%%%%%%%%
Two further tools combine static files (\ftfilestructure) and \ftversionclaim:
CMSmap\cite{cmsmap} (open source python CMS scanner) and
CMS-Garden CMSscanner\cite{cmsgardencmsscanner} (assumes you to have full access over the target machine).
%%%%%%%%%%%%%%%%%%%
droopescan\cite{droopescan} identifies issues with several content management systems, such as Drupal\cite{drupal} and Silverstripe\cite{silverstripe}.
They determine the target's software version by comparing md5-fingerprints of static files.

%\subsection{Other Tools}
A surplus amount of \emph{penetration tools} leverages various logical, security, or implementation faults to gain information about a system.
From this realm of penetration testing tools stems sqlmap \cite{sqlmap}.
Its main task consists of detecting and exploiting SQL flaws.
%, for example by injection, database server takeover, or data fetching.
It also comes with a fingerprint function for databases and the servers they are running on.
While the fingerprinting of the server is quite basic and relies on \ftversionclaim and \ftfilestructure, database versions can be determined using \ftversionspecific or \ftfunctionoutput.
This, of course, only works if the web application has an SQL injection flaw or the tool has direct access to the database.
%First, RFP supports further fingerprinting techniques like error message interpretation, dynamic function output comparison, dynamic request response comparison (a superset of dynamic boolean comparison), or non-desctructive security exploits.
%These allow RFP not only to adapt to a broad range of different software but it can also perform more fine grained scans.
%Second, sqlmap is (well) tailored for database management systems only, while RFP works for any given software in general (depending on the implementation), for example programming environments running PHP.
%Third, RFP is based on leveraging intrinsic properties of the software that is being analyzed.
%This approach results from the theoretical framework of RFP which, to the best of our knowledge, does not exist for sqlmap.

There are further tools on the market which have very few information about them, however, we want to list them for completeness.
The vulnerability scanner by Crashtest Security \cite{crashtestsecurity} has no public available information about their scanner, similar holds for the web application security test tools Rapid7 or AppSpider\cite{appspider}.
Retina CS\cite{retinacs} is a vulnerability scanner and management tool by BeyondTrust.
In comparison to \RFP, none of the tools presented in this section provide a theoretical framework or security analysis.

%\subsection{Academic Work}
\emph{Academic publications} regarding software fingerprinting are quite rare.
In \cite{DBLP:journals/popets/IrazoquiIES15}, the authors detect running crypto libraries in the cloud by exploiting leakages on the hardware level.% and also determine the IP of a co-located virtual machine.
Another approach to this field is software attestation, e.g., see \cite{DBLP:conf/ccs/ArmknechtSSW13} for an overview. 
Examples  are \cite{DBLP:conf/sp/SeshadriPDK04} for embedded devices, \cite{DBLP:journals/tifs/GardnerGR09} for voting machines, and \cite{DBLP:journals/adhoc/SeshadriLP11} for sensor networks.
The core idea is to run a challenge-response protocol over the code of the software and to detect cheating systems by measuring the time effort for a responses. 
This limits this approach to devices with restricted capabilities.
In particular, none of them is designed for web- or cloud applications.

\section{Conclusion}
Until now, as the comparison and analysis of employed fingerprinting techniques has shown (see Sections~\ref{sec:sota} and Appendix~\ref{sec:ftanalysis}), existing software trusts the provider or its service to output correct software version values, which may not be the actual version of the running software.
Therefore, it may occur that insecure software is believed to be secure.
We have presented \RFPlong which solves the \RSI problem, i.e., determining which version of a software is running remotely, e.g., at a cloud service provider.

To achieve this, a database is employed which contains challenges and expected replies of the target software, where each challenge depends on inherent, probabilistic parametric functionalities of the version that is tested.
Also, a strategy is utilized to efficiently choose the correct order of tests out of the database based on previous tests, producing a distinguishing sequence of challenge-response pairs based on intrinsic functions.
Furthermore, we introduce and leverage a theoretical framework based on finite state machines to built \RFP and yield its capabilities.
Formalization of the framework as well as testing software and version hierarchies are also a key part of \RFP.
This includes new fingerprinting techniques and an according security analysis of \RFP.
Furthermore, \RFP can be extended to work on hardware such as storage, but also to add an auditor to take away most of the burden of a regular user resulting in outsourced \RFP, i.e., \ORFP.

The \RFP scheme does not rely on the support of the service provider.
Since only the service or software itself is being tested, this enables direct verifiability of the running software.
Plus, \RFP can be applied independent of the software type, since intrinsic functions of the software are being leveraged and tested.
For example, it does not matter if \RFP tests forums, content management systems, programming languages, databases, or PaaS and SaaS in general.
To execute \RFP, two steps are required: (i) identification of characteristic functionalities of certain software versions and (ii) communicating with a software followed by an evaluation.
While (i) needs to be done manually, (ii) is performed automatically with a few queries to the target service.
This allows to efficiently test multiple systems at once in a very short amount of time.

While the properties of \RFP make it notably useful for the user, the CSP and auditor also gain benefits.
Auditors can automate and schedule security analyses, additionally, they may share or sell their developed databases and strategies.
Since \RFP is resistant against software manipulation of the CSP, the trust of customers will increase if an \RFP test has been performed successfully, e.g., by an auditor.
This leads to an increased security overall, since a malicious CSP risks to be debunked to not have updated its software.
For the CSP, this may result in an increasing number of customers.

The quality of an \RFP scheme depends on the employed database and the interfaces used therein.
Obviously, a software version can not be verified if the customer does not have access or a sufficiently populated database.
For the future, \RFP implementations for further software families are planned, e.g., Internet forums.
%A further open task is the creation of databases for any software.
Finally, it would be convenient to combine \RFP with software detection, i.e., detecting the software family.

In summary, \RFP reliably determines the actual software version $\swtarget$ running at a (cloud) service provider by employing a challenge-response protocol which requires the audited software $\swchallenge=\produce(\swsource)$ to perform randomized tasks based on intrinsic functionalities.
Since it is not economical for the CSP to implement simulation-hard functions of $\swtarget$ into $\swchallenge$ in order to pretend having installed $\swtarget$, \RFP yields based on a database of challenges, expected responses, time-constraints, and randomness parametrization the software version $\swtarget$ when auditing $\swchallenge$.
This finally allows the verifier to validate if $\swsource\equiv\swtarget$, i.e., if the service provider is honest.

\bibliographystyle{IEEEtran}
\bibliography{references}

\appendix

\section{Description of Fingerprint Techniques}\label{sec:ftdesc}
This is a thorough description of all employed fingerprinting techniques used by state of the art implementations of an \RSI scheme $\rsip$, see Section~\ref{sec:sota}.

\subsection{\ftclassone}
\subsubsection{\ftversionclaim}\label{sec:ftversionclaim}
The most simple and intuitive fingerprinting technique is to send a challenge with the task to reply with the current software version inside the running environment.
This information together with environmental data is sent as a response and evaluated, e.g., by extracting the version number using regular expressions.
Usually, this method is available for any kind of interface $\interface$, e.g., may even be public information.
For example, version number in the footer of a website or as part of a banner\footnote{Here, a banner means an ASCII rendering of text or an image with additional information like the version number and, e.g., the software authors.}.

One type of \ftversionclaim is known as the information disclosure attack.
In this attack, the remote software contains a usually publicly accessible resource that leaks internal information.
For web services, many PHP installation tutorials instruct the administrator to create a PHP file that calls the function \code{phpinfo()} for debugging purposes.
Various installed applications include such a file by default which can be accessed easily by a remote attacker, extracting a large amount of information about the software and server, e.g., PHP and operating system version, server IP address, configuration data, and so on.
Keep in mind that a malicious service provider can put arbitrary information in such a file, placing a red herring for attackers and fingerprinting tools at the same time.
See Remark~\ref{rem:regexhash} for a general description of an attack strategy for the software provider.

\subsubsection{\ftversionapi}\label{sec:ftversionapi}
The verifier will send a challenge to the API (application programming interface) which in turn directly yields the version number, hence no further processing is necessary.
For example, in PHP this can be realized by calling \code{phpinfo()} or in most SQL databases via \code{SELECT VERSION()}.
It is very similar to \ftversionclaim with the only difference that the employed challenge interface needs to have API access.
The reason why we include and differentiate this technique from \ftversionclaim is that in practice, either \ftversionclaim or a combination of both is used, increasing the capabilities of implementations.

\subsubsection{\ftversionspecific}\label{sec:ftversionspecific}
The idea of this fingerprinting technique is to challenge $\swchallenge$ to reply with a function that is specific for a single software version only, i.e., that literally depends on the software version.
Such a task consists of two steps: first, load the software version from storage, second, if the software version equals the specific version given in the challenge, perform a simple equation check of two randomly chosen equal numbers.
As a result, this will only be true, if the stored version number is the same as the version number contained in the challenge.
One example are MySQL queries which use comment execution\cite{mysqlcommentexecution}, i.e., a comments is interpreted and executed as a command only if the version requirement is fulfilled.
However, note that this technique can be reduced to \ftversionclaim due to reading the version label from storage.

% https://mariadb.com/kb/en/library/comment-syntax/
% https://github.com/MariaDB/server/blob/62bcd74712680fa07c9ed8c42c384c8825c4f9af/sql/sql_lex.cc#L1810

\subsubsection{\ftlistplugins}\label{sec:ftlistplugins}
This fingerprinting technique identifies the software version of a software $\sw$ by identifying the software versions of some ``sub-softwares'' $\tilde{\sw}:=\sw_1,\ldots,\sw_n$ that can be part of $\sw$, such as plug-ins.
In practice, implementations perform this test by using the techniques \ftversionclaim, \ftfilestructure (see below), or simply check for file existence for each $\sw\in\tilde{\sw}$.
Note that this technique requires a sufficient populated list $\tilde{\sw}$ to deduce $\version_\sw$.
For example, a content management system can load multiple plugins to enhance its functionality.
Then, each of them is placed in a certain subfolder which gets scanned by this fingerprinting technique for plugin versions or existence.

Note that this technique relies only on other techniques which are part of fingerprinting technique classes that are not secure, see Appendix~\ref{sec:ftanalysis}.
Even assuming these techniques would be secure, $\tilde{\sw}$ must be sufficiently large in order to conclude the version number of $\sw$, we currently see no software employing enough plugins in order to reliably determine its version.
Since this fingerprinting technique can be reduced to \ftversionclaim or \ftfilestructure, it belongs to both fingerprinting technique classes one and two.

\subsubsection{\fterrorcode}\label{sec:fterrorcode}
Similar to \ftversionclaim and \ftversionapi, $\verifier$ sends a query to \sw in order to retrieve a string containing the version number.
However, in this case, the goal is to provoke an error in the software to output an error message containing the version number.
For example, using a string as input for a function that is only defined for integer input.
One of the tools relying on this technique is nmap \cite{nmap}. incorporates this technique but can be defeated by appropriate tools \cite{fingerprintfucker,nmapdefeat,DBLP:conf/uss/SmartMJ00}.
Since this fingerprinting technique can be reduced to \ftversionclaim or \ftfilestructure, it belongs to both fingerprinting technique classes one and two.

If an error occurs in an underlying software $\sw$ that acts as a platform for $\swchallenge$, for example, a database to run a forum, information about $\sw$ can be retrieved through $\swchallenge$.
We are not aware that any software uses this technique but list it here for completeness, please see Remark~\ref{rem:llhl} for more details.

\subsection{\ftclasstwo}
\subsubsection{\ftfilestructure}\label{sec:ftfilestructure}
In this technique, a challenge of the verifier requests $\swchallenge$ to respond with a certain file.
The reply, i.e., a static file, is then parsed by the verifier, e.g., by using regular expressions or to check the structure of the file, e.g., order of configuration commands.
The outcome is compared to certain previously computed values which belong to a software version each.
For example, a verifier will search for certain text strings, HTML tag nesting, referenced URLs, simple key words, or the style of a readme file.
As for \ftversionclaim, a malicious provider can change the content of the static files, i.e., strings.
See Remark~\ref{rem:regexhash} for a general description of an attack strategy for the provider and Remark~\ref{rem:filerandomization} for a note on file randomization.

\subsubsection{\fthashdigest}\label{sec:fthashdigest}
This technique is sometimes referred to as ``aggressive verification'' by various implementations, since it checks every bit of a requested file.
A hash algorithm is applied to the file retrieved by a response and afterwards this value is compared to a pre-computed hash value stored at the verifier.
For example, the verifier may request a readme file and a configuration file.
On retrieval, he computes the hash value for each file and compares the output with his pre-computed values.
Note that this technique can be reduced to \ftfilestructure.

\subsection{\ftclassthree}
\subsubsection{\ftfunctionoutput}\label{sec:ftfunctionoutput}
The idea is to send challenges, where the according response depends on certain version-dependent functions to verify if these functions exist properly.
Parameters $p$ for a function $f$ can be fixed or randomized.
In essence, the challenges consists of performing the comparison $f(p)=f(p)$.
For example, sending a query to execute \code(content(var)=content(var)) with fixed values, will require the existence of the function \code{content} and the variable \code{var}.
However, this technique will always return a static binary value, i.e. \true or \false.

\subsection{\ftclassfour}
\subsubsection{\ftsecurityexploit}\label{sec:ftsecurityexploit}
This technique exploits a security vulnerability of the audited software.
If no update has been applied, the exploit will be successful.
For example, an authentication check may not be performed correctly, allowing access to otherwise protected information.

While vulnerability scanners test for security vulnerabilities, to the best of our knowledge they do not use the outcome to determine the version number.
Usually, first the version number is determined, e.g., by \ftversionapi, then one or more attacks are performed.
Note that this may lead to an abrupt halt of the software, which allows for no further tests or audits.

\begin{remark}[Regular Expression and Hash Digest Security]\label{rem:regexhash}
If $\rsip$ is open source (or reversible) and it uses regular expressions and/or a hash function as a fingerprinting technique, then this provides a blueprint to \prover how to pretend any version number.
The \prover only needs to investigate the regular expressions or possibly requested files and copy the required file from the software distributor's original repository (or similar source).
Then, he redirects any internal request to that file to a modified version of the file, while any external file access gets delivered the (original) file and hash value of the claimed version.
Also observe that if a honest \prover changes a file in a way that it contains negligible changes (e.g., white spaces) or indeed certain file changes like specific configurations or style changes, the software version did not change but the $\rsip$ employing the hash digest technique would yield a wrong result.
If \prover knows that a regular expression is used to find a certain value in a given file, he can simply insert a forged value for the certain expression such that the regular expression discovers the values the \prover wants to return.
If an $\rsip$ is not open source, \prover might learn over time which files are accessed, since usually only a very limited set (at most three) of specific files is part of $\rsip$.

In summary, this attack shows that neither regular expressions nor hash digests are suited for software version fingerprinting.
Using hash functions may even yield false negatives.
\end{remark}

\begin{remark}[Using Underlying Structures to Determine a Software Version]\label{rem:llhl}
Independent of software dependencies, only the software $\swchallenge$ is queried during $\rfps$.
In \RFP, we do not employ underlying or lower level software or structures (LL) as side-channel information for higher level software (HL).
Furthermore, a customer usually has no access to LL, but in the case he does, it is trivial to determine the version numbers of HL since he has full control thereof.

For the other way around, using HL to retrieve information about LL, the verifier could provoke errors in HL to produce an error message from LL, e.g., PHP running under Wordpress and an error in PHP due to a false used function of Wordpress.
However, this does not allow to precisely determine a version number of LL.
The reason is that the error message is static, can be changed to an arbitrary string, and will not depend on different functionalities of LL.
If HL is just a wrapper of LL, this comes down to testing LL directly by using another interface, for example, security flaws in HL allow accessing LL through HL.
Recall that $\rfps$ employs interfaces which can be used to address different access levels.
\end{remark}

\begin{remark}[On Randomization for File Structure and File Content]\label{rem:filerandomization}
To the best of our knowledge, in all current auditing softwares which rely on a technique based on static files such as \ftfilestructure and \fthashdigest, usually only a very limited set (in fact, at most three) of specific files is requested in the challenge.
Let us assume that any file of $\swchallenge$ can be chosen by the verifier and $\swchallenge$ provides enough files such that it is unfeasible for provider $\prover$ to create a file-switch for each file, i.e., decide which version to deliver depending on the verifier's challenge.
Then, randomization will not be economical, i.e., more expensive than running the agreed on software $\swtarget$, and hence out of interest for $\prover$.
\end{remark}

\section{Security Analysis of Fingerprinting Techniques}\label{sec:ftanalysis}
In this section we will provide a security analysis of the classes of fingerprinting techniques given in Section~\ref{sec:sota}.

\subsection{\ftclassone}
During setup, $\swtarget\in\swfamily$ is chosen and $\prover$ sets $\swchallenge = \produce(\swsource)$, where both softwares are identical except for $\version_\swchallenge \neq \version_\swsource$ and $\outputalphabet_\swchallenge = \outputs_\swsource\cup\version_\swchallenge$.
In the testing phase, $\verifier$ sends $\inp$ that requests $\swchallenge$ to output the value of the software version.
However, $\swchallenge$ returns $\outp \gets \version_\swchallenge$.
Finally, $\verifier$ outputs $\candidateset = \outp = \version_\swchallenge \neq \version_\swsource$, which contradicts $\rfpp$-soundness.
Also note that due to $\inp\not\in\dshard(\swtarget)$, it is immediately clear that the verifier can be fooled.

\subsection{\ftclasstwo}
During setup, $\swtarget\in\swfamily$ is chosen and $\prover$ sets $\swchallenge = \produce(\swsource)$, where both softwares are identical except for $\outputalphabet_\swchallenge = \bad{\outputalphabet}$ and $\fout_\swchallenge = \bad{\fout}$, such that $\bad{\fout}$ simulates a software with software version $\bad{\version}\in\bad{\outputalphabet}$ by yielding certain elements of $\bad{\outputalphabet}$.
Note that the value of $\version_\swchallenge$ does not matter, since the responses of $\swchallenge$ will not depend on it.
In the testing phase, $\verifier$ sends $\inp$ that requests $\swchallenge$ to output a (very small) set $\outputalphabet'_\swsource\subsetneq\outputalphabet_\swsource$.
However, $\swchallenge$ returns $\outp\gets\outputalphabet'_\swchallenge$.
Finally, $\verifier$ outputs $\candidateset = \bad{\version}\neq\version_\swsource$ derived from $(\inp, \outp)$, which contradicts $\rfpp$-soundness.
Also note that due to $\inp\not\in\dshard(\swtarget)$, it is immediately clear that the verifier can be fooled.

\subsection{\ftclassthree}
Let $\outputalphabet_\decision=\{\true,\false\}$.
During setup, $\verifier$ selects $\swtarget\in\swfamily$ such that it has the highest version for all softwares in $\swfamily$, i.e., $\version_\swtarget > \version_\sw \forAll \sw\in\swfamily$.
The $\prover$ sets $\swchallenge = \produce(\swsource)$, where both softwares are identical except for $\outputalphabet_\swchallenge = \bad{\outputalphabet}$ and $\ftrans_\swchallenge = \bad{\ftrans}$, where $\bad{\ftrans}$ simulates $\bad{\version}\in\bad{\outputalphabet}$.
Furthermore, he sets $\version_\swchallenge \neq \version_\swsource$ with $\{\version_\swchallenge\cup\outputalphabet_\decision\}\in\bad{\outputalphabet}$.
The testing phase is as follows:
\begin{enumerate}
\item $\verifier$ selects a function
\begin{align*}
\func \in \funcset_\swtarget\setminus\bigcup_{\substack{\sw\in\swfamily\\\version_\sw < \version_\swtarget}}\funcset_\sw
\end{align*}
and sends $\inp$ to $\swchallenge$, requesting the output of $\func$ which is defined for this fingerprinting technique as an element of $\outputalphabet_\decision$.
\item $\swchallenge$ returns $\outp \in \outputalphabet_\decision$.
\item If $\outp = \true$, $\verifier$ sets $\swtarget$ to be the software with the next lower version in $\swfamily$ and proceeds with step 2).\\
If $\outp = \false$, $\verifier$ stops.
\end{enumerate}
Since $\swchallenge$ computed each $\outp$ according to $\bad{\version}$, $\verifier$ outputs $\candidateset = \bad{\version}\neq\version_\swsource$ derived from all tuples $(\inp, \outp)$, which contradicts $\rfpp$-soundness.

Observe that $\swchallenge$ does not need to evaluate $\func$, he may output $\true$ and $\false$ according to any version $\bad{\version}$ he wants to simulate.
This especially holds if $\func$ is a simulation-hard function of $\swsource$, i.e., $\func\not\in\funcset_\swchallenge$.

Also note that due to $\inp\not\in\dshard(\swtarget)$ for all $\inp$ chosen in step 1), it is immediately clear that the verifier can be fooled.
We refer the reader to Remark~\ref{rem:secfunctionoutput} where we discuss the property $\rfps$-secure of the class 3 fingerprinting technique \ftfunctionoutput, also in respect to existing implementations.

\begin{remark}[$\rfps$ Secure \ftfunctionoutput]\label{rem:secfunctionoutput}
We describe and analyze the fingerprinting technique \ftfunctionoutput in Appendix~\ref{sec:ftfunctionoutput} and Appendix~\ref{sec:ftanalysis}, respectively.
To be secure according to $\rfps$, see Definition~\ref{def:rfps}, a fingerprinting technique needs also to be time-constrained, probabilistic, and employ intrinsic functions.
\ftfunctionoutput might indeed access intrinsic functions and a time-constrained can be added easily.
However, the response of $\swchallenge$ is always taken from the decision set $\{\true,\false\}$.
The probabilistic protocol $\rfpp$ requires the response to depend on the randomness of the challenge.
This is not given in a correct manner for this fingerprinting technique, since no matter how good the randomness source of the verifier is, the malicious prover has always a 50\% chance of correct guessing.
Note that the only existing implementation employing this fingerprinting technique\cite{sqlmap}, does not randomize the challenge.
Hence, \ftfunctionoutput is not $\rfps$-secure.
\end{remark}

\subsection{\ftclassfour}
This fingerprinting technique was actually never used to determine a version number, however, it might yield an error state ($\bot$), e.g., due to a deadlock or software crash.
In general, $\swchallenge$ might unintentionally halt and $\verifier$ can not run $\rfpp$ until the end.
Hence, $\rfpp$-correctness is not given.

\subsection{Overall Result}
The presented classes of fingerprinting techniques violate $\rfpp$-correctness or $\rfpp$-soundness and, hence, can be successfully fooled by a malicious prover $\prover$.

\section{Software Version Manipulation Script}\label{sec:vcas}
The bash script presented in Listing~\ref{lst:vcas} manipulates the software version of PHP by changing the configuration values before compiling.
The idea is that the strings defining the software version are configuration parameters of the installing process, and hence, may be changed nearly arbitrarily.
In Listing~\ref{lst:vcas}, these variables are set in Lines~3--7.
Note that there is also a ``real'' version variable which, in \RFP terminology, defines $\swsource$ while the fake ones change the software version string for $\swchallenge$.
In other words, this script essentially performs $\produce$ with $\swchallenge=\produce(\swsource)$.
See Section~\ref{sec:rsi} for more information about $\produce$.
Then, PHP 7.1.1 runs as PHP 20.9.85-car on the host system, e.g., Ubuntu 16.04.3 and Apache 2.

\begin{figure*}
\centering
\begin{minipage}{0.8\textwidth}
\begin{lstlisting}[basicstyle=\footnotesize, numbers=left, language=bash, frame=lines, framesep=5pt, caption={Software version manipulation script in bash for PHP 7.1.1 running on Ubuntu 16.04.3 and Apache 2.}, label=lst:vcas, breaklines=true]

echo "Fetch and install PHP"

PHP_VERSION_REAL="7.1.1"
PHP_VERSION_FAKE_MAJOR="20"
PHP_VERSION_FAKE_MINOR="9"
PHP_VERSION_FAKE_RELEASE="85"
PHP_VERSION_FAKE_EXTRA="-car"

cd /usr/src
if [[ `wget -S --spider "http://de2.php.net/get/php-$PHP_VERSION_REAL.tar.gz/from/this/mirror"  2>&1 | grep 'HTTP/1.1 200 OK'` ]]; then
    sudo wget -O "php-$PHP_VERSION_REAL.tar.gz" "http://de2.php.net/get/php-$PHP_VERSION_REAL.tar.gz/from/this/mirror"
else
    PHP_VERSION_REAL_MAJOR=${PHP_VERSION_REAL%%.*}
    sudo wget -O "php-$PHP_VERSION_REAL.tar.gz" "http://museum.php.net/php$PHP_VERSION_REAL_MAJOR/php-$PHP_VERSION_REAL.tar.gz"
fi
sudo tar -zxf "php-$PHP_VERSION_REAL.tar.gz"
sudo rm -rf "/usr/src/php-$PHP_VERSION_REAL.tar.gz"
sudo mv "php-$PHP_VERSION_REAL" "./php-$PHP_VERSION_FAKE"
cd "php-$PHP_VERSION_FAKE"
CONFIGFILE=$(ls configure.*) # file type string is not constant
sudo sed -i "/PHP_MAJOR_VERSION=/c\PHP_MAJOR_VERSION=$PHP_VERSION_FAKE_MAJOR" $CONFIGFILE
sudo sed -i "/PHP_MINOR_VERSION=/c\PHP_MINOR_VERSION=$PHP_VERSION_FAKE_MINOR" $CONFIGFILE
sudo sed -i "/PHP_RELEASE_VERSION=/c\PHP_RELEASE_VERSION=$PHP_VERSION_FAKE_RELEASE" $CONFIGFILE
sudo sed -i "/PHP_EXTRA_VERSION=/c\PHP_EXTRA_VERSION=$PHP_VERSION_FAKE_EXTRA" $CONFIGFILE
sudo ./buildconf --force
sudo ./configure "--prefix=$DIR_PHP" "--with-apxs2=$DIR_HTTPD/bin/apxs" "--with-config-file-path=$DIR_PHP" --with-mysql
sudo make
sudo make install
sudo libtool --finish ./libs
\end{lstlisting}
\end{minipage}
\end{figure*}

\section{Remark on Definition of $\decision$}
\begin{remark}[Definition of $\decision$]\label{rem:decision}
According to equation~\eqref{eq:f}, testing for a function $\func$ as part of $\rfpp$ will yield $\decision=\true$ for all $\sw$ with $\version_\swchallenge\geq\version_\sw$.
Let us get an overview of the different equality tests that can be employed here, i.e., $<$, $\leq$, $=$, $\geq$, and $>$.
For $=$ we have $\delta\gets\true$ iff $\version_\swchallenge = \version_\sw$.
However, equality is the overall goal of \RFP and can not be achieved by a single $\rfpp$ instance as explained in Section~\ref{sec:rfptest}.
Hence, this comparison does not make sense here.
For $\geq$ we have $\delta\gets\true$ iff $\version_\swchallenge\geq\version_\sw$, naturally represents the fact described in equation~\eqref{eq:f}.
Testing for $<$ yields if $\func$ has not yet been introduced to $\swfamily$, and complements the test for $\geq$ (and is reflected by $\decision\gets\false$ in $\rfpp$).
Using only $<$ would make testing for newest versions impossible.
Similar, a test for $\leq$ and $>$would also be suitable for $\rfpp$, representing checking for removed functions.
In practice, however, much more often functions will be added than removed, resulting in testing for $(\geq, >)$ being the best candidate.
\end{remark}

\section{Example Database Metadata}\label{sec:exampledbmeta}
In Listing~\ref{lst:dbmeta} we give an example for metadata configuration of a \RFP database for the software family PHP.

\begin{lstlisting}[basicstyle=\footnotesize, numbers=left, language=bash, frame=lines, framesep=10pt, numbersep=-8pt, caption={Example metadata of a \RFP database for PHP.}, label=lst:dbmeta]
  "creationTimestamp": "2018-10-17T03:32:29+02:00",
  "lastUpdateTimestamp": "2018-11-14T04:17:18+01:00",
  "defaultvalues": {
    "version.test.challenge.setstarttag": "true",
    "version.test.challenge.setendtag": "false",
    "version.test.expect.setstarttag": "false",
    "version.test.expect.setendtag": "false",
    "version.test.expect.type": "string",
    "version.test.label": "0",
    "version.test.variables.type": "rand",
    "version.test.variables.format": "value",
    "version.test.waittime.amount": 200,
    "version.test.waittime.type": "milliseconds"
  },
  "settings": {
    "interface.challenges": "ftp",
    "interface.responses": "http",
    "strategies": [
      "BinarySearch",
      "CascadingBinarySearch",
      "HighToLow",
      "LowToHigh",
      "MajorHighestStepUp"
    ]
  },
  "service": {
    "name": "php",
    "versions": { ... }
  }
\end{lstlisting}

\section{Related Work Comparison Table}\label{sec:relworktable}
In Table~\ref{tbl:relwork} we give an overview of related work in comparison of fingerprinting techniques, their classes, and relevant related techniques which possibly could be used as fingerprinting techniques, such as exploiting security flaws.

\begin{table*}
%\hspace*{-20mm}
%\begin{tiny}
\centering
\begin{tabular}{lcp{70mm}p{36mm}}
\toprule
\textbf{Implementation} & \textbf{FT Classes} & \textbf{Fingerprinting Techniques Employed} & \textbf{Relevant Related Techniques}\\
\midrule
\multicolumn{4}{c}{Penetration Frameworks}\\
Metasploit\cite{metasploit} & 1 & \ftversionclaim, \ftversionapi  & \ftsecurityexploit \\
Armitage\cite{armitage} & 1 & \ftversionclaim, \ftversionapi & \ftsecurityexploit \\
PenTesters Framework\cite{pentestersframework} & -- & -- & --\\
\midrule
\multicolumn{4}{c}{Vulnerability Scanners} \\
Burp Suite\cite{burp} & 1 & \ftversionclaim, \ftversionapi & \ftsecurityexploit \\
Nessus\cite{nessus} & 1 & \ftversionclaim, \ftversionapi & \ftsecurityexploit \\
Greenbone/OpenVAS\cite{openvas} & 1 & \ftversionclaim & \ftsecurityexploit \\
droopescan\cite{droopescan} & 1, 2 & \fthashdigest, \ftlistplugins & \ftsecurityexploit \\
nmap\cite{nmap} & 1 & \ftversionclaim, \ftversionapi, \fterrorcode (TCP) & -- \\
Software Version Reporter\cite{softwareversionreporter} & 1 & \ftversionclaim & -- \\
w3af\cite{w3af} & -- & -- & \ftsecurityexploit \\
Web Application Scanning\cite{webapplicationscanning} & 1 & \ftversionclaim, \ftversionapi & --\\
FreeScan\cite{freescan} & 1 & \ftversionclaim, \ftversionapi & --\\
Nexpose\cite{nexpose} & 1 & \ftversionapi & --\\
InsightVM\cite{insightvm} & 1 & \ftversionapi & --\\
OWASP Nettacker\cite{nettacker} & 1 & \ftversionapi & --\\
\midrule
\multicolumn{4}{c}{Online Version Scanners} \\
wappalyzer\cite{wappalyzer} & 1 & \ftversionclaim & --\\
WhatRuns\cite{whatruns} & 1 & \ftversionclaim & --\\
WhatWeb\cite{whatweb} & 1, 2 & \ftversionclaim, \fthashdigest & --\\
Guess\cite{guess}  & 1, 2 & \ftversionclaim, \ftfilestructure, \ftlistplugins & --\\
BlindElephant\cite{blindelephant} & 2 & \fthashdigest & --\\
WAFP\cite{wafp} & 2 & \fthashdigest & --\\
Static File Fingerprinting\cite{staticfilefingerprinting} & 2 & \fthashdigest & --\\
\midrule
\multicolumn{4}{c}{CMS Scanners} \\
What CMS?\cite{whatcms} & 1 & \ftversionclaim & --\\
WPScan\cite{wpscan} & 1, 2 & \ftversionclaim, \ftfilestructure & --\\
Plecost\cite{plecost} & 2 & \ftfilestructure & --\\
WPSeku\cite{wpseku} & 2 & \ftfilestructure & --\\
JoomScan\cite{joomscan} & 2 & \ftfilestructure & --\\
CMSScanner\cite{cmsscanner} & 2 & \ftfilestructure & --\\
CMSmap\cite{cmsmap} & 1, 2 & \ftversionclaim, \ftfilestructure & --\\
CMS-Garden CMSscanner\cite{cmsgardencmsscanner} & 1, 2 & \ftversionclaim, \ftfilestructure & --\\
\midrule
\multicolumn{4}{c}{Other Tools} \\
sqlmap\cite{sqlmap} & 1, 3 & \ftversionclaim, \fterrorcode, \ftversionspecific, \ftfunctionoutput & -- \\
\midrule
\multicolumn{4}{c}{This Work} \\
Reverse Fingerprinting & 5 & \fterrorhandling, \ftrequestresponse, \ftsecurityexploitnd & --\\
\bottomrule
\end{tabular}
\caption{Different tools and their employed fingerprinting techniques (FT).
Relevant related techniques represent FT that a tool does support, but are not used for fingerprinting.
FT class five is, in contrast to FT classes one to four, $\rfps$-secure.}
\label{tbl:relwork}
%\end{tiny}
\end{table*}

\end{document}